\documentclass[12pt]{article}
\usepackage{amsmath,amsthm}
\usepackage{graphicx}
\usepackage{enumerate}
\usepackage{natbib}
\usepackage{url} 
\RequirePackage[colorlinks,citecolor=blue,urlcolor=blue]{hyperref}
\usepackage{graphicx}
\usepackage{enumerate}
\usepackage{latexsym}
\usepackage{amssymb}
\usepackage{multirow}
\usepackage{float}
\usepackage{tikz}
\usetikzlibrary{decorations.pathreplacing,calligraphy}
\usepackage{bm}
\usepackage{paralist}
\usepackage{mathabx}
\usepackage{natbib}
\usepackage[ruled]{algorithm2e}
\usepackage{bigstrut,colortbl}
\usepackage{commath}
\usepackage{subcaption}
\usepackage{soul,xcolor}
\usepackage{subcaption}
\usepackage[inline]{enumitem}
\usepackage[figuresright]{rotating}
\usepackage[labelfont=bf]{caption}

\usepackage{thmtools}
\renewcommand\thmcontinues[1]{Continued}

\usepackage{pifont}
\usepackage{threeparttable}
{\end{quotation}}

\usepackage[normalem]{ulem}

\DeclareMathOperator*{\argmin}{argmin}
\renewcommand{\baselinestretch}{1.23}
\numberwithin{equation}{section}

\def \mcH{{\mathcal{H}_q}}
\def \mcX{\mathcal{X}}

\def \mcD{\mathcal{D}}

\def \mbI{\mathbb{I}}

\def \mbB{\mathbb{B}}

\def \mbS{\mathbb{S}}

\def \mbN{\mathbb{N}}
\renewcommand{\lg}{\langle}
\newcommand{\rg}{\rangle}

\def \cov{\textrm{cov}}

\def \bfx{\mathbf{x}}

\def \bfX{\mathbf{X}}
\def \bfx{\mathbf{x}}

\def \bfi{\mathbf{i}}
\def \bfj{\mathbf{j}}
\def \bfI{\mathbf{I}}
\def \bfbeta{\bm{\beta}}

\def \bfy{\mathbf{y}}

\def \bfY{\mathbf{Y}}

\def \ev{\mathbb{E}}
\def \var{\textrm{Var}}

\def \pr{\mathbb{P}}
\def \bfW{\mathbf{W}}

\def \bfR{\mathbf{R}}

\def \cip{\xrightarrow[\text{}]{\text{$\mathbb{P}$}}}

\newcommand{\vo}{\vec{o}\@ifnextchar{^}{\,}{}}

\newcommand\define{\mathrel{\stackrel{\makebox[0pt]{\mbox{\normalfont\tiny def}}}{=}}}

\theoremstyle{plain}

\newtheorem{theorem}{\indent Theorem}
\newtheorem*{theorem*}{\indent Theorem}
\newtheorem{corollary}{\indent Corollary}
\newtheorem{Assumption}{Assumption}

\newtheorem{lemma}{\indent Lemma}

\theoremstyle{definition}
\newtheorem{Example}{Example}
\DeclareSymbolFont{largesymbolsA}{U}{txexa}{m}{n}
\DeclareMathSymbol{\varprod}{\mathop}{largesymbolsA}{16}
\usepackage{xparse}

\graphicspath{{../}{./}}
\newcommand{\blind}{1}

\usepackage{multibib}
\newcites{App}{References}

\newcites{Supp}{Supplement References}

\graphicspath{{./}{figures/}}

\begin{document}

\def\spacingset#1{\renewcommand{\baselinestretch}%
{#1}\small\normalsize} \spacingset{1}


\if1\blind
{
  \title{\bf Estimation and Hypothesis Testing of Derivatives in Smoothing Spline ANOVA Models}
  \author{Ruiqi Liu
   \\
    Department of Mathematics and Statistics, Texas Tech University\\
            and \\
    Kexuan Li \\
    Global Analytics and Data Sciences, Biogen\\
        and \\
        Meng Li \\
    Department of Statistics, Rice University}
  \maketitle
} \fi

\if0\blind
{
  \bigskip
  \bigskip
  \bigskip
  \begin{center}
    {\LARGE\bf Estimation and Hypothesis Testing of Derivatives in Smoothing Spline ANOVA Models}
\end{center}
  \medskip
} \fi

\bigskip
\begin{abstract}
Within the framework of smoothing spline ANOVA, we propose a plug-in kernel ridge regression estimator to estimate the derivatives of the underlying multivariate regression function. We first establish an $L_\infty$ convergence rate of the proposed estimator under general random designs. When the covariates are uniformly distributed, we provide a in-depth analysis that includes a sharp upper bound and the minimax lower bound of the $L_2$ convergence rate. Additionally, motivated by a wide range of applications, we propose a hypothesis testing procedure to examine whether a derivative is zero. Theoretical results demonstrate that the proposed testing procedure achieves the correct size under the null hypothesis and is asymptotically powerful under local alternatives. 
For ease of use, we also develop an associated bootstrap algorithm to construct the rejection region and calculate p-value, and the consistency of the proposed algorithm is established. Simulation studies using synthetic data and an application to a real-world dataset confirm the effectiveness of our approach.
\end{abstract}

\noindent%
\vfill

\newpage
\spacingset{1.9} 
\section{Introduction}
Estimating derivatives of the underlying regression function is a important problem in nonparametric regression, with broad applications in various fields. For instance, in economics, derivatives play significant roles in determining partial marginal effects, measuring elasticities, and evaluating the curvature or concavity of functions (\citealp{charnes1985foundations,blundell1998semiparametric,shephard2015theory}). In manufacturing industry,  the structures of productivity curves such as local extrama can be used for abnormality detection (\citealp{woodall2004using}). Estimating the derivatives of the productivity curves can provide insight into identifying local extrema.   In the realm of deep learning, derivatives can be utilized to prune network structures (\citealp{hassibi1992second, dong2017learning}). In the context of spatial data analysis (\citealp{banerjee2003directional, meron2004vegetation, duhr2006molecules, chen2019pollution}), spatial gradients are often used to quantify the degree of change of a spatial surface at a given location or in a given direction.  Examples include meteorologists measuring temperature or rainfall gradients and environmental scientists evaluating  eco-environmental quality via pollution gradients. When it comes to statistics, bias-correction techniques that depend on estimated derivatives can be employed to achieve more reliable confidence intervals (\citealp{xia1998bias, calonico2018effect}).

Another important statistical problem is to examine whether a derivative is zero.  An example in economics is demand analysis (\citealp{deaton1986demand,labandeira2017meta,atalla2018gasoline}). A zero first-order derivative of the demand function with respect to price implies inelastic demand, while a zero second-order derivative  suggests a constant price elasticity. Additionally, researchers also noticed that  derivatives can serve as a tool for model selection (\citealp{gorsich2000variogram,YANG2003521}). For example, zero first-order partial derivatives suggests that it is safe to exclude the corresponding variables from the models, while the magnitude of some higher-order partial derivatives can help check if there are interactions among dependent variables. 

Although considerable progress has been made in estimating or inferring derivatives of univariate functions (e.g., see \citealp{wang2015derivative, liu2018derivative, liu2020smoothed}), the extension to multivariate functions is considerably more challenging due to various factors. One major hurdle is the curse of dimensionality, which arises from the fact that even a reasonably large number of data points may be inadequate with high-dimensional covariates, making it more difficult to obtain accurate estimates. As a result, high-dimensional problems generally have a much slower optimal convergence rate than their low-dimensional counterparts. Moreover, visualizing a general multivariate function and its derivatives is difficult and provides little insight into the effect of each covariate, which makes interpretation problematic.

The smoothing spline ANOVA (SS-ANOVA) model is a dimension-reduction framework for modeling multivariate functions and their derivatives (\citealp{lin2000tensor,WAHBA2003531,gu2013smoothing}).  This model  posits that the underlying regression function to be estimated can be expressed as a sum of one-dimensional functions (main effects), two-dimensional functions (two-way interactions), and so forth. The flexible structure provides an intuitive and efficient way in modeling the interactions among covariates and can be seen as an extension of additive models. More importantly, by controlling the orders of interactions, one can mitigate the issue of curse of dimensionality. Due to its advantages in modeling multivariate functions, the SS-ANOVA model has found widespread application across various areas including computer codes (\citealp{reich2009variable,touzani2013smoothing}), medical studies (\citealp{wahba1995smoothing,gao2001smoothing}), geographical sciences (\citealp{wahba1996smoothing,luo1998spatial,wang1998smoothing}), and complex data analysis (\citealp{zhang2018smoothing}).

\subsection{Related Works}
Kernel ridge regression (KRR) is a popular estimation method for SS-ANOVA models. 
The term ``Kernel Ridge Regression"  was firstly introduced in \cite{cristianini2000introduction} to describe a simplified variant of support vector regression. Due to its efficiency in dealing with nonlinear problems (e.g., see \citealp{hastie2009elements, gu2013smoothing}), KRR has gained increasing popularity and been widely applied in many areas including image processing (\citealp{an2007face, kumar2008face}), option pricing (\citealp{hu2020pricing}), forecasting (\citealp{haworth2014local, exterkate2016nonlinear}).

The analysis of KRR can be divided into two directions based on the purposes of applications. The first research line aims to quantify the estimation performance of the method, with existing works establishing bounds on the estimation error and convergence rates of the estimators. Along this line, \cite{zhang2005learning} derived generalization error bounds for KRR estimators in terms of the effective dimension, while \cite{steinwart2009optimal} obtained optimal generalization error bounds by assuming polynomial decay eigenvalues.  \cite{lin2000tensor}  proved a minimax optimal convergence rate of  KRR estimators in SS-ANOVA models, and  \cite{zhang2015divide} provided a non-asymptotic convergence analysis of KRR estimators with a general class of kernels. Moreover, KRR has also been applied to functional data  regression (\citealp{yuancai2010rkhs, cai2012minimax, sun2018optimal}) and quantile regression (\citealp{zhang2016quantile,lian2022distributed,wang2022sparse}).

The second direction of analysis focuses on applying KRR to statistical inference, with most existing works based on a Bahadur representation developed in \cite{shangejs2010}. For instance, \cite{sc13aos}  established uniform asymptotic inference results in smoothing spline models, which were further extended to semi-parametric models in  \cite{cs15}  and generalized functional linear models in \cite{shangaos2015}. More recently, the Bahadur representation has been applied to factor models (\citealp{zhao2021statistical}), semiparametric density models (\citealp{yu2022smoothing}), and functional linear quantile regression models (\citealp{sang2022statistical}).

The existing literature on kernel ridge regression (KRR) has primarily focused on estimating or conducting inference for the underlying regression function, rather than its derivatives. In contrast, derivative estimation has a long history in nonparametric statistics: sieve methods, local polynomial regression, and difference-based approaches have been extensively studied; see \cite{liu2020estimation} for a comprehensive review.

Works on derivative estimation via KRR remain relatively limited. \cite{dai2017minimax} established minimax optimal convergence rates for estimating both the regression function and its first-order partial derivatives in smoothing spline ANOVA (SS-ANOVA) models, under the additional assumption that observations of first-order partial derivatives are available. \cite{liu2020estimation} proposed a plug-in KRR estimator for partial derivatives of general orders and derived a nearly minimax-optimal rate when the covariates are uniformly distributed. More recently, \cite{TuoZou2024KRRFunctionals} developed asymptotic theory for linear functionals of KRR estimators; their results can be used to construct pointwise confidence intervals for derivatives of the regression function. However, their theory does not address global inferential questions, such as testing whether a derivative is identically zero over the entire domain.


\subsection{Contributions}
This research paper investigates SS-ANOVA models and introduces a plug-in KRR estimator, as well as a hypothesis testing approach, for estimating and inferring the partial derivatives of the multivariate regression function. Our main contributions can be summarized as follows.

\begin{enumerate}[label=(\alph*), ref=(\alph*)]
\item We establish an $L_\infty$ convergence rate for our proposed derivative estimator under general random designs, subject to the condition that the density of covariates is  bounded and bounded away from zero. Additionally, we provide an upper bound for the $L_2$ convergence when covariates are uniformly distributed. For partial derivatives with specific orders, we establish the minimax lower bound under $L_2$ norm, which matches our upper bound. These results are non-trivial extensions from previous works that focused on estimating the regression function (\citealp{lin2000tensor}) to now include its derivatives, and from univariate functions (\citealp{liu2020estimation}) to the multivariate case.  We note that \cite{TuoZou2024KRRFunctionals} study the asymptotic theory of KRR estimators in multivariate Sobolev spaces, which differ from the tensor product structure considered here.

\item    We develop a novel testing procedure for examining whether a partial derivative is zero. Prior testing procedures based on KRR estimators were primarily designed to test the function forms of underlying regression functions (e.g., see \citealp{cs15, liu2020distributed}). However, extending these methods to specification tests for derivatives is not trivial. The Wilks' type theorems (\citealp{sc13aos}) or uniform inference results (e.g., see \citealp{sc13aos, sang2022statistical}) in the existing literature rely on a Bahadur representation that demonstrates the influence function of the KRR estimator as a product of the noise term and an equivalent kernel. Furthermore, the equivalent kernel can be expressed as a linear combination of orthonormal eigenfunctions, which is essential in establishing the asymptotic distribution. However, the influence function of the derivative estimator does not have this nice property. As a result, the existing testing procedures cannot be directly adapted to our problem. To address this challenge, we construct a test statistic based on a maximum over numerous sampling points from the covariate domain. It is proved that the associated testing procedure has a correct size under the null hypothesis and a satisfactory power under local alternatives. To the best of our knowledge, this is the first work conducting hypothesis testing about derivatives in SS-ANOVA models.

\item Similar to the existing results in the literature (e.g., see \citealp{liu2020distributed}), the rejection region of the proposed testing procedure involves the eigenvalues and eigenfunctions of the reproducing kernel. However,  these unknown eigenpairs are often estimated by solving multiple integro-differential equations (e.g., see \citealp{cs15, sang2022statistical}), and its consistency remains unexplored. To address this limitation, we develop a theoretically valid bootstrap algorithm to construct the rejection region.  As shown in our simulation studies, the proposed bootstrap algorithm enjoys satisfactory finite sample performance.
\end{enumerate}

The rest of the paper is organized as follows. Section \ref{section:ssanova} reviews the preliminaries of SS-ANOVA models. In Section \ref{sec:derivative:estimation}, we propose plug-in KRR estimator for derivatives and investigate its convergence rate. Section \ref{section:testing} develops a hypothesis testing procedure based on the plug-in KRR estimator. An associated bootstrap algorithm is proposed to construct rejection region. Section
\ref{section:simulation} presents the results of Monte Carlo simulation, and Section \ref{section:empirical:application} provides the results of an empirical study. All the mathematical proofs are deferred to the Appendix.

\section{Smoothing Spline ANOVA}\label{section:ssanova}
In this section, we review the SS-ANOVA model in the nonparametric regression problem.
Suppose observations $\mcD_n=\{(\bfX_i, Y_i), i=1,\ldots, n\}$ are generated from the following model
\begin{eqnarray*}
Y=f_0(\bfX)+\epsilon,
\end{eqnarray*}
where $X\in \mcX\define [0, 1]^r$ is the covariate vector with density $\pi_{\bfX}$, $Y\in \mathbb{R}$ is the response, and $\epsilon$ is the unobserved error. Moreover, we assume that the underlying regression function $f_0$ admits the following decomposition:
\begin{eqnarray}
f_0(\bfx)&=&\textrm{constant}+\sum_{s=1}^r g_s(x_s)+\sum_{1\leq s_1<s_2\leq r}g_{s_1,s_2}(x_{s_1}, x_{s_2})+\ldots \nonumber \\
&&+\sum_{1\leq s_1<\ldots<s_q\leq r}g_{s_1,\ldots, s_q}(x_{s_1},\ldots, x_{s_q}).\label{eq:f0:decomposition}
\end{eqnarray}
Here $\bfx=(x_1,\ldots, x_r)^\top \in \mcX$. The component $g_{s_1,\ldots, s_k}$ models the interaction among $x_{s_1},\ldots, x_{s_k}$, with an interaction order of $k$.  The highest order of interactions in (\ref{eq:f0:decomposition}) is $q\in \{1,\ldots, r\}$. The flexible structure of $f_0$  allows us to control the order of interactions. Specifically, in many applications, we have prior knowledge that $q$, the highest order of interactions, is strictly less than $r$. For instance, a widely used setting is $q=1$, which means there are no interactions among the covariates, and it reduces to an additive model. In this paper, we are interested in estimating the underlying regression function $f_0$ and its derivatives.

In nonparametric regression problems, it is common to impose some smoothness assumptions on  $f_0$. When $\bfX$ is univariate, i.e. $r=1$, a widely used function class is the Sobolev-Hilbert space of univariate functions defined as
\begin{eqnarray*}
\mbS_m\define \left\{f: \int_0^1 \left|f^{(k)}(x)\right|^2 dx<\infty, \textrm{ for all } k=0,1,\ldots, m \right\},
\end{eqnarray*}
where $m\geq 0$ is an integer specifying the degree of smoothness. The Sobolev inner product for this class is $\lg f, g\rg_{\mbS_m}=\sum_{k=0}^{m-1}\int_0^1\int_0^1 f^{(k)}(x)g^{(k)}(y)dxdy  +\int_0^1 f^{(m)}(x)g^{(m)}(x)dx$\footnote{Using Poincar\'{e} inequality, it is easy to see that the norm $\|f\|_{\mbS_m}^2$ is equivalent to the standard Sobolev norm $\sum_{k=0}^m \int_0^1 |f^{(k)}(x)|^2dx$.}.

When $r>1$, the tensor product space provides an intuitive way in modeling the multivariate regression functions. Specifically, for Hilbert spaces $H_s$ of functions of $x_s$ with $s=1,\ldots, r$,  the tensor product space $\otimes_{s=1}^r H_s$ is the completion of the class $\{\sum_{i=1}^k  \prod_{s=1}^r f_{si}(x_s):  f_{si}\in H_s, k\geq 1\}$.
Here the completion is under an inner product $\lg\cdot, \cdot\rg_{\otimes_{s=1}^r H_s}$ that satisfies $\lg  \prod_{s=1}^r f_{s}, \prod_{s=1}^r g_{s} \rg_{\otimes_{s=1}^r H_s}=\prod_{s=1}^r\lg f_s, g_g\rg_{H_s}.$  A popular choice (e.g., \citealp{lin2000tensor, gu2013smoothing}) is  $\otimes_{s=1}^r \mathbb{S}_m$, the tensor product space with $H_1=\ldots=H_r=\mathbb{S}_m$. The product structure of $\otimes_{s=1}^r \mathbb{S}_m$ is flexible in modeling  interactions among covariates. To see this, let us decompose $\mathbb{S}_m$ into a direct sum of two orthogonal subspaces,
\begin{eqnarray*}
\mathbb{S}_m=\{1\}\oplus \mathbb{S}_m^{(0)},
\end{eqnarray*}
where $\{1\}$ is the space of constants, and $\mathbb{S}_m^{(0)}=\{f\in \mathbb{S}_m: \int_0^1 f(x)dx=0\}$. Consequently, it yields the decomposition
\begin{eqnarray}
\otimes_{s=1}^r \mathbb{S}_m(x_s)&=&\otimes_{s=1}^r \left(\{1\}\oplus \mathbb{S}_m^{(0)}(x_s)\right)\nonumber\\
&=&\{1\}\oplus\left\{\sum_{s=1}^r S_m^{(0)}(x_s)\right\}\oplus\left\{\sum_{1\leq s_1<s_2\leq r} S_m^{(0)}(x_{s_1})\otimes S_m^{(0)}(x_{s_2})\right\}\oplus\ldots\nonumber\\
&&\oplus\ldots \oplus \left\{ S_m^{(0)}(x_{1})\otimes \ldots \otimes S_m^{(0)}(x_{r})\right\}. \label{eq:space:decomposition}
\end{eqnarray}
where we use $x_i$'s to denote the arguments in the corresponding spaces. For any $f\in \otimes_{s=1}^r \mathbb{S}_m$, (\ref{eq:space:decomposition}) implies the following decomposition
\begin{eqnarray}
f(\bfx)=\textrm{constant}+\sum_{s=1}^r g_s(x_s)+\sum_{1\leq s_1<s_2\leq r}g_{s_1,s_2}(x_{s_1}, x_{s_2})+\ldots +g_{1,\ldots, r}(x_1,\ldots, x_r),\label{eq:component:f:decomposition}
\end{eqnarray}
with $\bfx=(x_1,\ldots, x_r)^\top\in \mcX$.  Similar to (\ref{eq:f0:decomposition}), the component $g_{s_1,\ldots, s_k}$ here models the interaction of the corresponding covariates. However, the highest order of interactions  in (\ref{eq:component:f:decomposition}) is $r$, which is a special case of (\ref{eq:f0:decomposition})  with $q=r$. To make it more flexible and allow for general $q\in \{1,\ldots, r\}$, we consider using the following space:
\begin{eqnarray*}
\mcH&\define&\{1\}\oplus\left\{\sum_{s=1}^r S_m^{(0)}(x_s)\right\}\oplus\left\{\sum_{1\leq s_1<s_2\leq r} S_m^{(0)}(x_{s_1})\otimes S_m^{(0)}(x_{s_2})\right\}\\
&&\oplus\ldots \oplus \left\{\sum_{1\leq s_1<\ldots<s_q\leq r} S_m^{(0)}(x_{s_1})\otimes \ldots \otimes S_m^{(0)}(x_{s_q})\right\}.
\end{eqnarray*} 
Clearly, $\mcH$ is a subspace of $\otimes_{s=1}^r \mathbb{S}_m$, and its inner product satisfies $\lg f, g\rg_\mcH=\lg f, g\rg_{\otimes_{s=1}^r \mathbb{S}_m}$ for all $f,g\in \mcH$. More importantly, all the functions in $\mcH$ admit the decomposition in (\ref{eq:f0:decomposition}). Consequently, we assume $f_0\in \mcH$  throughout this paper. 

To end this section, we introduce some commonly used notations. For positive sequences $a_n$ and $b_n$, we say $a_n\lesssim (\gtrsim) b_n$ if $a_n\leq (\geq) Cb_n$ for some $C>0$ and for all $n$ large enough. We say $a_n\asymp b_n$ if $a_n\lesssim b_n$ and $a_n\gtrsim b_n$.  For a sequence of random vectors $X_n$ and a sequence of deterministic scalars $a_n>0$, we say $X_n=O_P(a_n)$ if $\lim_{C\to \infty}\lim_{n\to \infty}\pr(\|X_n\|>Ca_n)=0$.  We write $X_n=o_P(a_n)$ if $\lim_{n\to \infty}\pr(\|X_n\|>\delta a_n)=0$ for any $\delta>0$. Here $\|\cdot\|$ is the Euclidean norm of vectors. The notation $\cip$ stands for convergence in probability. 
For a multivariate function $f:\mcX \to \mathbb{R}$, we use $\|f\|_{\sup}$ to denote its supremum norm and $\|f\|_{L_2}=\sqrt{\int_{\mcX} f^2(\bfx)\pi_{\bfX}(\bfx)d\bfx}$ to denote the $L_2$ norm. Moreover, for any vector $\bfbeta=(\beta_1,\ldots, \beta_r)^\top$ with $\beta_i$'s being non-negative integers, we use $\partial^{\bfbeta} f$ to denote the partial derivative $\partial^{|\bfbeta|}f/\partial x_1^{\beta_1}\ldots \partial x_r^{\beta_r}$, where $x_1,\ldots, x_r$ are the arguments and $|\bfbeta|=\sum_{i=1}^r \beta_i$. Finally, we use $\mbN_+$ to represent the collection of all  positive integers and define $\mbN=\mbN_+\cup \{0\}$. For any multi-index $\bfi=(i_1,\ldots, i_r)$ and vector  $\bfbeta=(\beta_1,\ldots, \beta_r)^\top$, we define $\bfi^{\bfbeta}=i_1^{\beta_1}\ldots i_r^{\beta_r}$.

\section{Derivative Estimation} \label{sec:derivative:estimation}

\subsection{Plug-in Estimator}
In SS-ANOVA models, the classical KRR estimates $f_0$ by solving the following minimization problem:
\begin{eqnarray*}
\widehat{f}\define \argmin_{f\in \mcH} \left\{\frac{1}{2n}\sum_{i=1}^n\left(Y_i-f(\bfX_i)\right)^2+\frac{\lambda}{2} \|f\|^2_\mcH \right\},
\end{eqnarray*} 
where $\lambda> 0$ is the tuning parameter. 

The above is an optimization problem over a space of functions, and a popular tool to solve this problem is  the  reproducing kernel hilbert space (RKHS). It is well known that $\mbS_m$ is a RKHS endowed with the inner product $\lg\cdot, \cdot \rg_{\mbS_m}$. Specifically, let 
\begin{eqnarray*}
R(x, y)=1+\sum_{v=1}^{m} \frac{B_v(x)B_v(y)}{(v!)^2}+(-1)^{m-2}\frac{B_{2m}(|x-y|)}{(2m)!}
\end{eqnarray*}
for $x,y\in [0,1]$ with $B_v$'s being the Bernoulli polynomials. It can be verified that $R(x, y)$ is the reproducing kernel for   $\mbS_m$ that satisfies  $\lg f, R_x \rg_{\mbS_m}=f(x)$ for every $f\in \mbS_m$; see \cite{gu2013smoothing}. Here $R_x(\cdot)=R(x, \cdot)$. It is worth mentioning that $\widetilde{R}(x,y)\define R(x, y)-1$ is the reproducing kernel for the subspace $\mbS_m^{(0)}$.

Extension from $\mbS_m$ to $\mcH$ can be easily made using the standard kernel construction procedure. To proceed, let us define 
\begin{eqnarray*}
R^{(q)}(\bfx, \bfy)&=&1+\sum_{s=1}^r \widetilde{R}(x_s, y_s)+\sum_{1\leq s_1<s_2\leq r} \widetilde{R}(x_{s_1}, y_{s_1})\widetilde{R}(x_{s_2}, y_{s_2})\\
&&+\ldots+\sum_{1\leq s_1<\ldots<s_q\leq r} \widetilde{R}(x_{s_1}, y_{s_1})\ldots \widetilde{R}(x_{s_q}, y_{s_q}),
\end{eqnarray*}
for $\bfx=(x_1,\ldots, x_r)^\top, \bfy=(y_1,\ldots, y_r)^\top \in \mcX$. It was shown in \cite{gu2013smoothing} that $R^{(q)}(\bfx, \bfy)$ is a reproducing kernel of $\mcH$ under the inner product $\lg \cdot, \cdot \rg_\mcH$. In a special case when $q=r$, a simple expression of the reproducing kernel is $R^{(r)}(\bfx, \bfy)=\prod_{s=1}^r R(x_s, y_s)$.

Relying on the above notations and the representer theorem (e.g., see \citealp{gu2013smoothing}), the explicit formula of the estimator is given by
\begin{eqnarray}
\widehat{f}(\bfx)=\Phi^\top(\bfx)\left(\bfR+n\lambda \bfI\right)^{-1}\bfY, \label{eq:solution:hat:f}
\end{eqnarray}
where $\Phi(\bfx)=(R^{(q)}(\bfx, \bfX_1),\ldots, R^{(q)}(\bfx, \bfX_n))^\top \in \mathbb{R}^n$, $\bfY=(Y_1,\ldots, Y_n)^\top \in \mathbb{R}^n$, and $\bfR=[R^{(q)}(\bfX_i, \bfX_j)]\in \mathbb{R}^{n\times n}$. According to the plug-in principle, a natural estimator of $\partial^{\bfbeta} f_0$ is 
\begin{eqnarray}
\partial^{\bfbeta}\widehat{f}(\bfx)=\left(\partial^{\bfbeta}\Phi(\bfx)\right)^\top\left(\bfR+n\lambda \bfI\right)^{-1}\bfY, \label{eq:solution:hat:f:derivative}
\end{eqnarray}
where $\partial^{\bfbeta}\Phi(\bfx)$ is the entry-wise $\bfbeta$-directional derivative of $\Phi(\bfx)$. 

Since $\widehat{f}, f_0\in \mcH$, the highest order of interaction for both functions is $q$.  It follows that $\partial^{\bfbeta} \widehat{f}=\partial^{\bfbeta} f_0=0$ if $\sum_{i=1}^r I(\beta_i>0)>q$. Therefore, we can restrict our attention to the case when $\bfbeta\in \mbB_q\define \{(\beta_1,\ldots, \beta_r)^\top \in \mathbb{N}^r: \sum_{i=1}^r I(\beta_i>0)\leq q \}$. 

\subsection{Convergence Analysis}\label{section:convergence:analysis}
Before proceeding, let us review the eigensystem in $\mcH$. Let $\mbI_r=\{\bfi=(i_1,\ldots, i_r): i_1,\ldots, i_r\in \mbN_+\}$ and $\mbI_q=\{\bfi \in \mbI_r: \sum_{k=1}^r I(i_k>1)\leq q\}$. From \cite{lin2000tensor}, we see that if  $\pi_{\bfX}$, the density function $\bfX$, is uniformly bounded and bounded away from zero, then there is a sequence of eigenvalues $\rho_{\bfi}$ and eigenfunctions $\psi_{\bfi}$ with $\bfi \in \mbI_q$ such that
\begin{eqnarray*}
\lg \psi_{\bfi}, \psi_{\bfj} \rg_{L_2}=\delta_{\bfi \bfj},\quad \lg \psi_{\bfi}, \psi_{\bfj} \rg_{\mcH}=\delta_{\bfi \bfj}/\rho_\bfi,
\end{eqnarray*}
where $\delta_{\bfi \bfj}$ is the Kronecker delta. Moreover, the eigenvalue satisfies $\rho_\bfi \asymp i_1^{-2m}\ldots i_r^{-2m}$, and the eigenfunctions $\psi_{\bfi}$'s form an orthonormal basis of $\mcH$ in terms of $\lg \cdot, \cdot \rg_{L_2}$.   Using the above notations, we introduce two quantities closely related to the convergence rate of the plug-in estimator:
\begin{eqnarray*}
h\define \left(\sum_{\bfi\in \mbI_q} \frac{1}{1+\lambda/\rho_\bfi}\right)^{-1}, \quad h_{\bfbeta}\define \left(\sum_{\bfi\in \mbI_q} \frac{\bfi^{\bfbeta}}{1+\lambda/\rho_\bfi}\right)^{-1}.
\end{eqnarray*}
Here $h^{-1}$ is known as the effective dimension in the literature (see \citealp{zhang2005learning}), which is crucial in the convergence rate of $\widehat{f}$. Noting that $h=h_{\bfbeta}$ with $\bfbeta=0$, the quantity $h_{\bfbeta}^{-1}$ essentially plays a similar role in the convergence analysis of the derivative estimator $\partial^{\bfbeta}\widehat{f}$.

The following assumptions are needed to establish convergence rate of the proposed estimator.
\begin{Assumption}\label{Assumption:distribution}
\makeatletter
\hyper@anchor{\@currentHref}%
\makeatother
\begin{enumerate}[label=(\roman*), ref=(\roman*)]
\item  \label{A1:density} There is a constant $C_\pi\geq 1$ such that $C_\pi^{-1}\leq \pi_\bfX(\bfx)\leq C_\pi$ for all $\bfx\in \mcX$.
\item \label{A1:moment:error} The noise term $\epsilon$ is sub-Gaussian, i.e., $\pr(|\epsilon|>t)\leq 2\exp\{-t^2/(2\sigma^2)\}$ for some $\sigma>0$ and for all $t>0$. Moreover, for some constant $C_\sigma>0$, it holds that $\ev(\epsilon|\bfX)=0$,
 $\ev(\epsilon^2|\bfX)\geq C_\sigma^{-1}$ and $\ev(\epsilon^k | \bfX)\leq C_\sigma^k$ for $k=1,2,3,4$.
\end{enumerate}
\end{Assumption}
\begin{Assumption}\label{Assumption:eigen}
It holds that
$\sup_{\bfx\in \mcX}|\psi_\bfi(\bfx)|\leq C_\psi$ and $\sup_{\bfx\in \mcX}|\partial^{\bfbeta}\psi_\bfi(\bfx)| \leq C_\psi \bfi^{\bfbeta}$ for all $\bfi \in \mbI_q$ and for some constant $C_\psi>0$.
\end{Assumption}
Assumption \ref{Assumption:distribution}\ref{A1:density} is the standard assumption on nonparametric regression problems; see \cite{h98,sc13aos,lin2000tensor,liu2020distributed}. Under this assumption, \cite{lin2000tensor} showed that $\rho_\bfi\asymp i_1^{-2m}\ldots i_r^{-2m}$, which is crucial in the asymptotic analysis. Assumption \ref{Assumption:distribution}\ref{A1:moment:error} is the distributional assumption on the noise term. Specifically, it requires that $\epsilon$ is sub-Gaussian with zero conditional mean. Moreover, $\epsilon$  also has bounded  conditional moments, and its conditional variance is bounded away from zero. 

Assumption \ref{Assumption:eigen} imposes some boundedness conditions on the eigenfunctions, and similar assumptions were also imposed in \cite{sc13aos, liu2020estimation, liu2020distributed, zhao2021statistical}.   For general density $\pi_\bfX$, it is difficult to verify Assumption \ref{Assumption:eigen}. However, when $\pi_\bfX$ is the uniform distribution over $\mcX$, the expression of $\psi_{\bfi}$ is available, and Assumption \ref{Assumption:eigen} will be satisfied. The following lemma summarizes this fact.
\begin{lemma}\label{lemma:eigen:function:uniform:design}
If $\pi_\bfX$ is the uniform distribution over $\mcX$, then 
\begin{eqnarray}
\psi_{\bfi}(\bfx)=\varphi_{i_1}(x_1)\varphi_{i_2}(x_2)\ldots \varphi_{i_r}(x_r), \label{eq:cos:sin:basis}
\end{eqnarray}
where $\varphi_1(x)=1, \varphi_{2i}(x)=\sqrt{2}\sin(2i\pi x), \varphi_{2i+1}(x)=\sqrt{2}\cos(2i\pi x)$  for $i\in \mathbb{N}$. 
\end{lemma}

\begin{theorem}\label{theorem:convergence:rate}
Under Assumptions \ref{Assumption:distribution} and \ref{Assumption:eigen}, if  $nh^2\to \infty$, $\lambda \to 0$, and $\max_{1\leq i \leq r}\beta_i\leq m-1$, then
\begin{eqnarray*}
\|\partial^{\bfbeta}\widehat{f}-\partial^{\bfbeta}f_0\|_{\sup}^2=O_P\left(\frac{\lambda}{h_{2\bfbeta}}+\frac{1}{nhh_{2\bfbeta}}\right).
\end{eqnarray*}
\end{theorem}

Theorem \ref{theorem:convergence:rate} provides an $L_\infty$ convergence rate of the plug-in estimator $\partial^{\bfbeta}\widehat{f}$. However, since the rate involves $\lambda, h_{2\bfbeta}, h$, and $n$, it is not convenient for conducting in-depth analysis. Using the following lemma, we are able to express $h_{\bfbeta}$ and $h$  in terms of $\lambda$, which significantly simplifies the presentation.
\begin{lemma}\label{lemma:rate:h:beta}
Under Assumptions \ref{Assumption:distribution} and \ref{Assumption:eigen}, if $2m>\beta_{\max}+1$ and $\lambda \to 0$, then we have
\begin{eqnarray*}
h_{\bfbeta}^{-1}\asymp \lambda^{-\frac{\beta_{\max} +1}{2m}}[\log(1/\lambda)]^{N_{\max}\wedge q-1},
\end{eqnarray*}
where $\beta_{\max}=\max_{1\leq i\leq r}\beta_i$ and $N_{\max}=\sum_{i=1}^r I(\beta_i=\beta_{\max})$. As a consequence, it follows that $h^{-1}\asymp \lambda^{-\frac{1}{2m}}[\log(1/\lambda)]^{q-1}$.
\end{lemma}
Lemma \ref{lemma:rate:h:beta} is an important result in this paper, based on which we can quantify the role of $\lambda$ played in the convergence rate. Compared with the existing results in the literature, Lemma \ref{lemma:rate:h:beta} is essentially a non-trivial generalization from $\bfbeta=0$ to $\bfbeta\neq 0$. For example, when $\bfbeta=0$ and $q=1$, it is well known that $h^{-1}\asymp \lambda^{-\frac{1}{2m}}$ (e.g., see \citealp{liu2020estimation, liu2020distributed}). In addition,  \cite{lin2000tensor} proved that $h^{-1}\asymp \lambda^{-\frac{1}{2m}}[\log(1/\lambda)]^{q-1}$ with $\bfbeta=0$. As a comparison, Lemma \ref{lemma:rate:h:beta} allows for general $\bfbeta$ and $q$. 

Combining Theorem \ref{theorem:convergence:rate} with Lemma \ref{lemma:rate:h:beta}, we lead to the  following corollary.
\begin{corollary}\label{corollary:convergence:rate}
Under Assumptions \ref{Assumption:distribution} and \ref{Assumption:eigen}, if $n\lambda^{1/m}[\log(n)]^{2-2q}\to \infty$, $\lambda \to 0$, and $\beta_{\max}\leq m-1$, then 
\begin{eqnarray}
\|\partial^{\bfbeta}\widehat{f}-\partial^{\bfbeta}f_0\|_{\sup}^2=O_P\left(\lambda^{1-\frac{2\beta_{\max}+1}{2m}}[\log(n)]^{N_{\max}\wedge q-1}+n^{-1}\lambda^{-\frac{\beta_{\max}+1}{m}}[\log(n)]^{N_{\max}\wedge q+q-2} \right).\label{eq:corollary:convergence:rate:eq:1}
\end{eqnarray}
As a consequence, we have the following statements:
\begin{enumerate}[label=(\roman*), ref=(\roman*)]
\item \label{corollary:convergence:rate:item:1} If $\lambda\asymp \left(n[\log(n)]^{1-q}\right)^{-2m/(2m+1)}$, then 
\begin{eqnarray*}
\|\partial^{\bfbeta}\widehat{f}-\partial^{\bfbeta}f_0\|_{\sup}^2=O_P\left(n^{-\frac{2m-2\beta_{\max}-1}{2m+1}} [\log(n)]^{\frac{(q-1)(2m-2\beta_{\max}-1)}{2m+1}+N_{\max}\wedge q-1}\right);
\end{eqnarray*}
\item \label{corollary:convergence:rate:item:2} If $\lambda\asymp n^{-\nu}$ with $0<\nu< m/(\beta_{\max}+1) $, then $\|\partial^{\bfbeta}\widehat{f}-\partial^{\bfbeta}f_0\|_{\sup}=o_P(1)$.
\end{enumerate}
\end{corollary}
The convergence rate presented in Corollary \ref{corollary:convergence:rate} provides a bias and variance decomposition. To see this, note that $\beta_{\max}\leq m-1$ implies that $1>(2\beta_{\max}+1)/(2m)$. Consequently, the first term on the right side of (\ref{eq:corollary:convergence:rate:eq:1}) increases as $\lambda$ becomes larger, representing the bias. Meanwhile, the second term on the right side of (\ref{eq:corollary:convergence:rate:eq:1}) corresponds to the variance, as it decreases when increasing $\lambda$. As shown in Statement \ref{corollary:convergence:rate:item:1}, their sum can be minimized through a particular selection of $\lambda$. Furthermore, when $\lambda$ is appropriately chosen, Statement \ref{corollary:convergence:rate:item:2} reveals the consistency of the plug-in estimator in terms of $L_\infty$ norm.

\subsection{Minimax Optimal Rate}\label{section:lower:bound}
The convergence rate obtained in the previous section is not minimax optimal under a general density $\pi_\bfX$. In this section, we will explore the performance of the plug-in estimator when the $\bfX_i$'s are uniformly distributed.

We first present a theorem providing the $L_2$ convergence rate of the plug-in estimator with general choices of $\bfbeta=(\beta_1, \ldots, \beta_r)^\top \in \mbB_q$.
\begin{theorem}\label{theorem:convergence:rate:uniform:design}
Under the conditions of Theorem \ref{theorem:convergence:rate}, if $\pi_{\bfX}$ is the uniform distribution over $\mcX$, then 
\begin{eqnarray*}
\|\partial^{\bfbeta} \widehat{f}-\partial^{\bfbeta}f_0\|_{L_2}^2=O_P\left(\lambda^{1-\frac{\beta_{\max}}{m}}+n^{-1}\lambda^{-\frac{1+2\beta_{\max}}{2m}}[\log(n)]^{q-1} \right).
\end{eqnarray*}
In particular, when $\lambda\asymp \left(n[\log(n)]^{1-q}\right)^{-2m/(2m+1)}$, it follows that
\begin{eqnarray*}
\|\partial^{\bfbeta} \widehat{f}-\partial^{\bfbeta}f_0\|_{L_2}^2=O_P\left\{\left(\frac{n}{[\log(n)]^{q-1}}\right)^{-\frac{2(m-\beta_{\max})}{2m+1}}\right\}.
\end{eqnarray*}
\end{theorem}
Compared to Corollary \ref{corollary:convergence:rate}, Theorem \ref{theorem:convergence:rate:uniform:design} yields an improved convergence rate of $n^{-\frac{1}{2m+1}}$ (up to a logarithmic factor). The reason behind this improvement is that when $\bfX_i$'s are uniformly distributed, the eigenfunctions have explicit expressions, which are provided in (\ref{eq:cos:sin:basis}). A desirable theoretical property of this sequence of eigenfunctions is that $\partial^{\bfbeta}\psi_{\bfi}$'s remain orthogonal under $\lg \cdot, \cdot \rg_{L_2}$. This preservation of orthogonality enables us to analyze the $L_2$ convergence and obtain a sharper upper bound. A direct consequence of Theorem \ref{theorem:convergence:rate:uniform:design} is the following corollary.
\begin{corollary}\label{corollary:convergence:rate:uniform:design}
Assume that $\bfbeta=(\beta_1,\ldots, \beta_r)^\top \in \mbB_q$ satisfies one of the following conditions:
\begin{enumerate}[label=(\roman*), ref=(\roman*)]
\item $\beta_1=\ldots=\beta_r=0$;
\item \label{corollary:convergence:rate:uniform:design:type:2} $\beta_1, \ldots, \beta_r\in \{0, \beta\}$ and $\sum_{i=1}^r I(\beta_i>0)=q$ for some integer $\beta\in [1, m-1]$.
\end{enumerate}
Moreover, if   the conditions of Theorem \ref{theorem:convergence:rate:uniform:design} hold and $\lambda\asymp \left(n[\log(n)]^{1-q}\right)^{-2m/(2m+1)}$, then
\begin{eqnarray*}
\|\partial^{\bfbeta} \widehat{f}-\partial^{\bfbeta}f_0\|_{L_2}^2=O_P\left\{\left(\frac{n}{[\log(n)]^{q-1}}\right)^{-\frac{2(m-\beta)}{2m+1}}\right\}.
\end{eqnarray*}
\end{corollary}
Corollary \ref{corollary:convergence:rate:uniform:design} considers the estimation of two special types of derivatives. The first type is the regression function $f_0$ itself, which corresponds to $\bfbeta=0$. The second type considers partial derivatives of the form $\partial^{q\beta}f_0/(\partial x_{i_1}^\beta \ldots \partial x_{i_q}^\beta)$, for some $\beta\in [1, m-1]$ and $i_1,\ldots, i_q \in \{1,\ldots, r\}$. Derivatives of this type are also important in many statistical applications such as model selection, which are illustrated by  Examples \ref{example:1}-\ref{example:3}.

\begin{Example}\label{example:1}
When $\beta=q=1$, the underlying regression function becomes 
\begin{eqnarray}
f_0(\bfx)=\textrm{constant}+\sum_{s=1}^r g_s(x_s).\label{eq:example:1}
\end{eqnarray}
Hence, derivatives satisfying \ref{corollary:convergence:rate:uniform:design:type:2} in Corollary \ref{corollary:convergence:rate:uniform:design} should be $\partial^{\bfbeta} f_0=g_s'(x_s)$ for some $s=1,\ldots, r$. The magnitude of $g_s'(x_s)$ helps determine if $x_s$ should be included in the model.
\end{Example}

\begin{Example}\label{example:2}
When $\beta=2$ and $q=1$, the underlying regression function is the same as (\ref{eq:example:1}).
Accordingly,  derivatives satisfying \ref{corollary:convergence:rate:uniform:design:type:2} in  Corollary \ref{corollary:convergence:rate:uniform:design} become $\partial^{\bfbeta} f_0=g_s''(x_s)$ for some $s=1,\ldots, r$. The magnitude of $g_s''(x_s)$ tells if $g_s$ can be simplified to a linear function.
\end{Example}

\begin{Example}\label{example:3}
When $\beta=1$ and $q=2$, the underlying regression function becomes 
\begin{eqnarray*}
f_0(\bfx)=\textrm{constant}+\sum_{s=1}^r g_s(x_s)+\sum_{1\leq i<j\leq r} g_{i,j}(x_{i}, x_{j}).
\end{eqnarray*}
Hence, derivatives satisfying \ref{corollary:convergence:rate:uniform:design:type:2}  in Corollary \ref{corollary:convergence:rate:uniform:design} are of the form $\partial^{\bfbeta} f_0=\partial^2 g_{i, j}/ \partial x_{i} \partial x_{j}$ for some $1\leq i<j\leq r$. If $\partial^2 g_{i, j}/ \partial x_{i} \partial x_{j}=0$ for all $1\leq i<j\leq r$, then we can remove all the interaction terms, and $f_0$ is reduced to (\ref{eq:example:1}).
\end{Example}

Compared with existing literature, Corollary \ref{corollary:convergence:rate:uniform:design} generalizes results in \cite{liu2020estimation} from univariate regression ($q=r=1$) to multivariate cases ($r>1$). Furthermore, it improves upon the rate $[\log(n)/n]^{-2(m-\beta)/(2m+1)}$ in \cite{liu2020estimation} by eliminating the $\log(n)$ factor when $q=r=1$. It is also noteworthy that the rate in Corollary \ref{corollary:convergence:rate:uniform:design} is minimax optimal, as confirmed by the following theorem.

\begin{theorem}\label{theorem:lower:bound:equal:beta}
Let $\Omega$ be the collection of all possible joint distributions of $(\bfX^\top,Y)^\top$. Moreover, for some constant $C>0$, let us define
\begin{eqnarray*}
\Omega_m=\{\pr_f\in \Omega:  Y=f(\bfX)+\epsilon,   f\in \mcH, \|f\|_\mcH \leq  C, \epsilon \textrm{ and } \bfX \textrm{ satisfy Assumption \ref{Assumption:distribution}} \}.
\end{eqnarray*}
If $\bfbeta=(\beta_1,\ldots, \beta_r)^\top \in \mbB_q$ satisfies one of the following conditions:
\begin{enumerate}[label=(\roman*), ref=(\roman*)]
\item $\beta_1=\ldots=\beta_r=0$;
\item $\beta_1, \ldots, \beta_r\in \{0, \beta\}$ and $\sum_{i=1}^r I(\beta_i>0)=q$ for some integer $\beta\in [1, m-1]$;
\end{enumerate}
then there is a constant $c>0$ not relying on $n$ such that
\begin{eqnarray*}
\inf_{\widehat{f}}\sup_{\pr_f\in \Omega_m}\ev_{\pr_f}\left(\left\|\partial^{\bfbeta}\widehat{f}-\partial^{\bfbeta}{f}\right\|_{L_2}^2\right)\geq c\left(\frac{n}{[\log(n)]^{q-1}}\right)^{-\frac{2(m-\beta)}{2m+1}}.
\end{eqnarray*}
Here $\ev_{\pr_f}$ is the expectation associated with the distribution $\pr_f$, and the infimum is taking over all estimators  based on $n$ observations.
\end{theorem}
Theorem \ref{theorem:lower:bound:equal:beta} provides the minimax lower bound for estimating $\partial^{\bfbeta}f_0$ in SS-ANOVA models when $\bfbeta$ satisfies certain structures. When $q=r=1$, the lower bound in Theorem \ref{theorem:lower:bound:equal:beta} matches the result in \cite{stone1982optimal}. When $\beta=0$, it also coincides with the minimax rate of estimating $f_0$ in \cite{lin2000tensor}. Thus, Theorem \ref{theorem:lower:bound:equal:beta} generalizes the univariate case to the multivariate case and extends from $\beta=0$ to general $\beta\geq 0$.
\section{Hypothesis Testing} \label{section:testing}
\subsection{Testing Procedure}\label{section:testing:procedure}
In many statistical applications, it is often of interest to test the following hypotheses:
\begin{eqnarray*}
H_0: \partial^{\bfbeta} f_0=0 \textrm{ V.S. } H_1: \partial^{\bfbeta} f_0\neq 0.
\end{eqnarray*}
For example, $\partial f_0/\partial x_1=0$ suggests that we can remove $x_1$ from the model, which can serve as a tool for variable selection. Moreover, $\partial^2 f_0/\partial x_1 \partial x_2=0$ implies no interactions between $x_1$ and $x_2$. In this section, we propose a testing procedure based on the plug-in estimator $\partial^{\bfbeta}\widehat{f}$ to examine $H_0$. Before proceeding, we first introduce the concept of equivalent kernel.

Given smoothing parameter $\lambda>0$, the equivalent kernel $K(\cdot, \cdot): \mcX\times \mcX \to \mathbb{R}$ is a function such that 
\begin{eqnarray}
K(\bfx, \bfy)\define \sum_{\bfi\in \mbI_q} \frac{\psi_{\bfi}(\bfx)\psi_{\bfi}(\bfy)}{1+\lambda/\rho_\bfi}, \label{eq:equivalent:kernel}
\end{eqnarray}
where $(\psi_{\bfi}, \rho_\bfi)$'s are the eigenpairs introduced in Section \ref{section:convergence:analysis}. We write $K_{\bfx}(\cdot)=K(\bfx, \cdot)$, and let $\partial^{\bfbeta}K_{\bfx}(\widetilde{\bfx})$ denote the derivative of $\widetilde{\bfx}\to K_{\bfx}(\widetilde{\bfx})$.  Using the above notation, our procedure can be summarized in three steps.
\begin{enumerate}[label=(\roman*), ref=(\roman*)]
\item Generate $p$ i.i.d. observations $\bfx_1,\ldots, \bfx_p$ from a prespecified density $\omega(\bfx)$ over $\mcX$, and construct the statistic $\max_{1\leq i\leq p}\sqrt{nh_{2\bfbeta}}|\partial^{\bfbeta}\widehat{f}(\bfx_i)|$.
\item Let  $G_P(g_\bfx)$ be a Gaussian process indexed by $\bfx \in \mcX$ with covariance function $\cov(g_\bfx(\zeta),g_{\widetilde{\bfx}}(\zeta))$, where $g_{\bfx}(\zeta)=\sqrt{h_{2\bfbeta}} \partial^{\bfbeta} K_\bfX(\bfx)\epsilon$ and $\zeta=(\epsilon, \bfX^\top)^\top$. Obtain the critical value $q_{\alpha, n}$ such that $\pr(\max_{1\leq i\leq p}|G_P(g_{\bfx_i})|> q_{\alpha, n})=\alpha.$
\item Reject $H_0$ if $\max_{1\leq i\leq p}\sqrt{nh_{2\bfbeta}}|\partial^{\bfbeta}\widehat{f}(\bfx_i)|> q_{\alpha, n}$.
\end{enumerate}

The following additional assumptions are required to investigate the performance of the proposed testing procedure.
\begin{Assumption}\label{Assumption:asymptotic:normal}
It holds that  $\inf_{\bfx\in \mcX} h_{2\bfbeta}\sum_{\bfi \in \mbI_q}{ |\partial^{\bfbeta}\psi_{\bfi}(\bfx)|^2}{(1+\lambda/\rho_{\bfi})^{-2}}\geq c_\psi$ for some constant $c_\psi>0$. 
\end{Assumption}
\begin{Assumption}\label{Assumption:sampling:density}
It holds that $C_\omega^{-1}\leq \omega(\bfx)\leq C_\omega$ for some $C_\omega>1$ and all $\bfx\in \mcX$.
\end{Assumption}

Assumption \ref{Assumption:asymptotic:normal} is a technical assumption. Essentially,  the variance of $G_P(g_{\bfx})$ is represented by the term $h_{2\bfbeta}\sum_{\bfi \in \mbI_q}{ |\partial^{\bfbeta}\psi_{\bfi}(\bfx)|^2}{(1+\lambda/\rho_{\bfi})^{-2}}$. The purpose of its uniform lower bound is to enable the application of the anti-concentration inequality for the maximum of Gaussian random variables (see \citealp{chernozhukov2016empirical}). A similar assumption was also used in \cite{sc13aos} for the case where $\bfbeta=0$.

Assumption \ref{Assumption:sampling:density} states that the sampling density $\omega(\bfx)$ must be uniformly bounded and bounded away from zero. An example is the uniform distribution over $\mcX$.

\begin{theorem}\label{theorem:asymptotic:testing}
Under Assumptions \ref{Assumption:distribution}-\ref{Assumption:sampling:density}, if $\lambda n^{m}[\log(n)]^{-m(3q-N_{\max}\wedge q-1)}\to \infty$, $\lambda n \log(n)\to 0$, and $\beta_{\max}\leq m-1$, then the following statements hold.
\begin{enumerate}[label=(\roman*), ref=(\roman*)]
\item Under $H_0: \partial^{\bfbeta}f_0=0$, it follows that 
\begin{eqnarray*}
\sup_{t\in \mathbb{R}}\left|\pr\left(\max_{1\leq i\leq p} \sqrt{nh_{2\bfbeta}}|\partial^{\bfbeta} \widehat{f}(\bfx_i)|\leq t\right)-\pr\left(\max_{1\leq i\leq p}|G_P(g_{\bfx_i})|\leq t\right)\right|\to 0.
\end{eqnarray*}
As a consequence, it follows that
\begin{eqnarray*}
\lim_{n\to \infty}\pr\left(\max_{1\leq i\leq p} \sqrt{nh_{2\bfbeta}}|\partial^{\bfbeta} \widehat{f}(\bfx_i)|>q_{\alpha, n}\right)=\alpha.
\end{eqnarray*}
\item Let $d_n=(nh_{2\bfbeta})^{-1/2}\log(n)$. If $d_n\sqrt{p}\to \infty$ and $\omega(\bfx)\in [C_\omega^{-1}, C_\omega]$ for some $C_\omega>1$ and for all $\bfx\in \mcX$, then for any sequence $C_n\to \infty$, it holds that
\begin{eqnarray}
\lim_{n\to \infty}\inf_{f_0\in \Sigma_{C_nd_n}}\pr_{f_0}\left(\max_{1\leq i\leq p} \sqrt{nh_{2\bfbeta}}|\partial^{\bfbeta} \widehat{f}(\bfx_i)|>q_{\alpha,n}\right)=1,\nonumber 
\end{eqnarray}
where $\Sigma_{C_nd_n}=\{f_0\in \mcH: \|\partial^{\bfbeta}f_0\|_{L_2}\geq C_n d_n, \|\partial^{\bfbeta}f_0\|_{\sup}\leq C_{\bfbeta}\}$ with $C_{\bfbeta}<\infty$ being some constant, and $\pr_{f_0}$ indicates that the observations $(\bfX_i^\top, Y_i)^\top$'s are generated with the regression function $f_0$. Consequently, if $\lambda\asymp n^{-1}[\log(n)]^{-2}$, then it holds that
\begin{eqnarray*}
d_n\asymp n^{-\frac{2m-2\beta_{\max}-1}{4m}}[\log(n)]^{\frac{2\beta_{\max}+1}{2m}+ \frac{N_{\max}\wedge q+1}{2}}.
\end{eqnarray*}
\end{enumerate}
\end{theorem}

The first statement of Theorem \ref{theorem:asymptotic:testing} indicates that the proposed testing procedure has an asymptotic size of $\alpha$. The second statement, on the other hand, states that the testing procedure is capable of rejecting the $H_0$ when the underlying regression function satisfies $\|\partial^{\bfbeta} f_0\|_{L_2}\gtrsim d_n$. With a suitable choice of $\lambda$, it follows that $d_n\asymp n^{-\frac{2m-2\beta{\max}-1}{4m}}$ (up to a logarithm factor). The proof of Theorem \ref{theorem:asymptotic:testing} relies on the following asymptotic expansion:
\begin{eqnarray}
\sqrt{nh_{2\bfbeta}}\|\partial^{\bfbeta}\widehat{f}-\partial^{\bfbeta}e-\partial^{\bfbeta}f_0\|_{\sup}=o_P(1),\nonumber
\end{eqnarray}
where $e(\bfx)=n^{-1}\sum_{i=1}^n \epsilon_i K_{\bfX_i}(\bfx)$ for $\bfx\in \mcX$, $\partial^{\bfbeta} e=n^{-1}\sum_{i=1}^n \epsilon_i \partial^{\bfbeta} K_{\bfX_i}$, and $K_{\bfX_i}(\cdot)=K(\bfX_i, \cdot)$ and $\partial^{\bfbeta} K_{\bfX_i}$ are the equivalent kernel defined in (\ref{eq:equivalent:kernel}) and its derivative.

The testing procedure is closely related to but significantly distinct from the existing works. When $\bfbeta=0$, the hypothesis testing procedures proposed in \cite{liu2020distributed, pmlr-v99-liu19a} are based on $\|\widehat{f}\|_{L_2}^2\approx \|e\|_{L_2}^2$ under $H_0$, while \cite{sc13aos} and \cite{ sang2022statistical} established weak convergence of the process $\bfx \to e(\bfx)$. All these existing results make use of the orthonormality of $\psi_\bfi$, which plays a crucial role in their analysis. However, when $\bfbeta\neq 0$, it becomes challenging to prove asymptotic normality of $\|\partial e\|^2_{L_2}$ or weak convergence of the process $\bfx \to \partial^{\bfbeta} e(\bfx)$, as the sequence $\partial^{\bfbeta} \psi_{\bfi}$'s is no longer orthogonal. This motivates our testing procedure that approximates the maximum over $\mcX$ using the maximum over the sampling points $\{\bfx_1,\ldots, \bfx_p\}$, while the limiting distribution of the latter can be established by the coupling techniques in \cite{chernozhukov2016empirical}.

\subsection{Bootstrap Algorithm}\label{section:bootstrap}
The testing procedure proposed in the previous section enjoys nice theoretical properties. However, computing both $h_{2\bfbeta}$ and $q_{\alpha, n}$ makes its implementation impractical, as it requires calculating the eigenpairs $(\rho_\bfi, \psi_\bfi)'$s. To address this limitation, we introduce a bootstrap algorithm that can automatically construct the rejection region and calculate the p-value. Before proceeding, let us define the bootstrap estimator of $f_0$ as follows
\begin{eqnarray}
		\widehat{f}^*\define \argmin_{f\in \mcH} \left\{\frac{1}{2n}\sum_{i=1}^n W_i\left(Y_i-f(\bfX_i)\right)^2+\frac{\lambda}{2} \|f\|^2_\mcH \right\}.\label{eq:bootstrap:estimator}
\end{eqnarray}
Here   $W_1,\ldots, W_n$ are i.i.d. nonnegative random weights that satisfy Assumption \ref{Assumption:bootstrap} below. Similar to  (\ref{eq:solution:hat:f}),  the solution can be explicitly calculated as
\begin{eqnarray*}
\widehat{f}^*(\bfx)=\Phi^\top(\bfx)\left(\bfW\bfR+n\lambda \bfI\right)^{-1}\bfW\bfY,
\end{eqnarray*}
where $\bfW\in \mathbb{R}^{n\times n}$ is the diagonal matrix with diagonal elements $W_1,\ldots, W_n$. The corresponding bootstrap plug-in estimator  is 
\begin{eqnarray}
\partial^{\bfbeta} \widehat{f}^*(\bfx)=\left(\partial^{\bfbeta} \Phi(\bfx)\right)^\top\left(\bfW\bfR+n\lambda \bfI\right)^{-1}\bfW\bfY.\label{eq:solution:hat:f:star:derivative}
\end{eqnarray}

\begin{Assumption}\label{Assumption:bootstrap}
The weight $W_i$ is nonnegative that satisfies $\ev(W_i)=\var(W_i)=1$ and $\pr(W_i>t)\leq C_We^{-t/C_W}$ for some $C_W>0$ and for all $t>0$.
\end{Assumption}
The requirement of nonnegative weights in Assumption \ref{Assumption:bootstrap} is to ensure the convexity of the optimization problem in (\ref{eq:bootstrap:estimator}). Furthermore, the tail probability condition on $W_i$ is mild and can be satisfied by many distributions. For instance, if $P(W_i=0)=P(W_i=2)=1/2$, it corresponds to the Rademacher weights. Another common choice for $W_i$ is the exponential distribution with mean one.

\begin{theorem}\label{theorem:bootstrap:asymptotic}
Under Assumptions \ref{Assumption:distribution}-\ref{Assumption:bootstrap}, if $\lambda n^{m} [\log(n)]^{-2m(q+7)}\to \infty$, $\lambda \to 0$, and $\beta_{\max}\leq m-1$, then it follows that
\begin{eqnarray*}
\sup_{t\in \mathbb{R}}\left|\pr\left(\max_{1\leq i\leq p} \sqrt{nh_{2\bfbeta}}|\partial^{\bfbeta} \widehat{f}^*(\bfx_i)-\partial^{\bfbeta} \widehat{f}(\bfx_i)|\leq t\bigg| \mcD_n\right)-\pr\left(\max_{1\leq i\leq p}|G_P(g_{\bfx_i})|\leq t\right)\right|\cip 0,
\end{eqnarray*}
where $G_P$ is the Gaussian process defined in Section \ref{section:testing:procedure}, and $\mcD_n=\{(\bfX_i^\top, Y_i)^\top\}_{i=1}^n$ are the observations. As a consequence, it follows that
\begin{eqnarray*}
\pr\left(\max_{1\leq i\leq p} \sqrt{nh_{2\bfbeta}}|\partial^{\bfbeta} \widehat{f}^*(\bfx_i)-\partial^{\bfbeta} \widehat{f}(\bfx_i)|>q_{\alpha,n}\bigg| \mcD_n\right)\cip \alpha.
\end{eqnarray*}
\end{theorem}
Theorem \ref{theorem:bootstrap:asymptotic} shows that the conditional distribution of  $\eta_n^*\define \max_{1\leq i\leq p} \sqrt{nh_{2\bfbeta}}|\partial^{\bfbeta} \widehat{f}^*(\bfx_i)-\partial^{\bfbeta} \widehat{f}(\bfx_i)|$ is asymptotically equivalent to $\max_{1\leq i\leq p}|G_P(g_{\bfx_i})|$ in Kolmogorov distance either under $H_0$ or $H_1$. A direct implication of Theorems  \ref{theorem:asymptotic:testing} and \ref{theorem:bootstrap:asymptotic} is that we can use the empirical quantile of $\eta_{n, (1)}^*, \ldots, \eta_{n, (B)}^*$ to estimate $q_{\alpha, n}$, as well as to estimate the p-value. Here $\eta_{n, (k)}^*$'s is a bootstrap sample of $\eta_n^*$ with a sample size $B$. This idea is formally summarized by Algorithm \ref{alg:bootstrap}. Additionally, utilizing the scale-invariance property of sample quantiles,  Algorithm \ref{alg:bootstrap} does not require estimating $h_{2\bfbeta}$, which makes it more practically convenient.

\begin{algorithm}[ht!]
\caption{Bootstrap  Algorithm}\label{alg:bootstrap}
\SetKwInOut{Input}{Input}
\SetKwInOut{Output}{Output}
\KwIn{Data $(\bfX_i, Y_i), i=1,\ldots, n$; Smoothing parameter $\lambda>0$; Sampling density $\omega(\bfx)$; Bootstrap sample size: $B\in \mbN_+$; Bootstrap weight distribution $P_W$; Significant level $\alpha \in(0, 1);$}

		Generate i.i.d. observations $\bfx_1,\ldots, \bfx_p$ from the density $\omega(\bfx)$\;
		Use  (\ref{eq:solution:hat:f:derivative}) to calculate the  estimator $\partial^{\bfbeta}\widehat{f}$;
		
		Calculate the unscaled statistic $\phi_n=\max_{1\leq i\leq  p}|\partial^{\bfbeta}\widehat{f}(\bfx_i)|$;
		
\For{$k=1$ \KwTo $B$}
	{   
		Generate i.i.d. random variables $W_1^{(k)}, \ldots W_n^{(k)}$ from distribution $P_W$\;
		
		Calculate $\bfW^{(k)}=\textrm{Diag}(W_1^{(k)}, \ldots W_n^{(k)})$;

		Calculate the bootstrap estimator $\partial^{\bfbeta}\widehat{f}^*_{(k)}$  by replacing 	 $\bfW$ with $\bfW^{(k)}$ in (\ref{eq:solution:hat:f:star:derivative});
	
		Generate i.i.d. observations $\bfx_1^{(k)},\ldots, \bfx_p^{(k)}$ from the density $\omega(\bfx)$\;
		Construct the bootstrap sample $\phi^*_{n,(k)}=\max_{1\leq i\leq  p}|\partial^{\bfbeta}\widehat{f}^*_{(k)}(\bfx_i^{(k)})-\partial^{\bfbeta}\widehat{f}(\bfx_i^{(k)})|$.
	}
Let $t_{\alpha, n}$ be the upper $\alpha$-quantile of $\phi^*_{n,(1)},\ldots, \phi^*_{n,(B)}$\;


\KwOut{Rejct $H_0$ if and only if $\phi_n>t_{\alpha, n}$, p-value $\sum_{k=1}^B I(\phi^*_{n,(k)}\geq \phi_n)/B$.}
\end{algorithm}

\section{Monte Carlo Simulation}\label{section:simulation}
In this section, we conduct extensive simulation studies to examine the finite sample performance of the proposed method on synthetic datasets.  For all experiments, we consider Sobolev-Hilbert space $\mbS_m$ with the degree of smoothness $m=2$. Given parameter $b\in \mathbb{R}$, three  data generation processes (DGP) are considered:
\begin{itemize}
    \item DGP 1: $f_0(x_1, x_2) = f_{01}(x_1) + f_{02}(x_2) + bf_{03}(x_1,x_2)$, where $f_{01}(x_1)=\exp\{-4(1-2x_1^2)\}(1-2x_1)$, $f_{02}(x_2)= \sin(8x_2) + \cos(8x_2) + \log(4/3 + x_2)$, $f_{03}(x_1, x_2)= 1.5\exp(x_1+x_2)$.
In this DGP, we use the gradient direction $\bfbeta = (1, 1)^\top$.

\item DGP 2: $f_0(x_1, x_2, x_3, x_4, x_5) =bf_{01}(x_1) + f_{02}(x_2) + f_{03}(x_3) + f_{04}(x_4) + f_{05}(x_5)$, where $f_{01}(x_1)=5 x_1$, $f_{02}(x_2)=3(2 x_2-1)^2$, $f_{03}(x_3)=4 \sin (2 \pi x_3) /(2-\sin (2 \pi x_3))$, $f_{04}(x_4) = 2x_4^3 + \min(x_4, 0.2) + \max(x_4, 0.8)$, and $f_{05}(x_5)=6(0.1 \sin (2 \pi x_5)+0.2 \cos (2 \pi x_5)+0.3 \sin (2 \pi x_5)^2+0.4 \cos (2 \pi x_5)^3+0.5 \sin (2 \pi x_5)^3)$. 
The gradient direction  is set to $\bfbeta = (1, 0, 0, 0, 0)^\top$.

\item DGP 3: $f_0(x_1, x_2, x_3) = b\left((f_{01}(x_1) + f_{012}(x_1, x_2) + f_{013}(x_1, x_3)+ f_{0123}(x_1, x_2, x_3)\right)
 + f_{02}(x_2) + f_{03}(x_3) + f_{023}(x_2, x_3)$, where $f_{01}(x_1)=x_1$, $f_{02}(x_2)=(2x_2-1)^2$, $f_{03}(x_3)=\exp(x_3-0.5)$, $f_{012}(x_1, x_2)=x_1\sin(x_2)$, $f_{023}(x_2, x_3) = x_2^2x_3$, $f_{013}(x_1, x_3)=x_3\sin(x_1)$, and $f_{0123}(x_1, x_2, x_3)=\frac{x_1}{x_2 + x_3}$. 
The gradient direction is assigned as $\bfbeta = (1, 0, 0)^\top$.
\end{itemize}

Int each DGP, we use the error distribution $\epsilon \sim N(0, \sigma^2)$ with $\sigma$ controlling the level of noise, while the signal strength is controlled by the parameter $b$. In particular, $b=0$ is used to examine the empirical size of the proposed test under $H_0: \partial^{\bfbeta} f_0=0$,  and $b\neq 0$ is to examine the empirical power of the proposed test. To implement Algorithm \ref{alg:bootstrap},  the bootstrap sample size is set as $B=500$, and the target significance level is chosen as $\alpha=0.1$. The sampling density $w$ and weight distribution $P_W$ are uniform and $Exp(1)$, respectively. Following \cite{liu2026optimal} and others,  we choose the regularization parameter $\lambda$  by maximizing the (pseudo) marginal likelihood $\bfY \mid \bfX \sim N\left(0, \hat{\sigma}_n^2(n \lambda)^{-1} \bfR+\hat{\sigma}_n^2 \bfI\right)$,  where $\hat{\sigma}_n^2\define \lambda \bfY^T\left[\bfR+n \lambda \bfI\right]^{-1} \bfY$.  
\subsection{Estimation}
\begin{figure}[ht!]
  \centering
 \includegraphics[width=1.97 in]{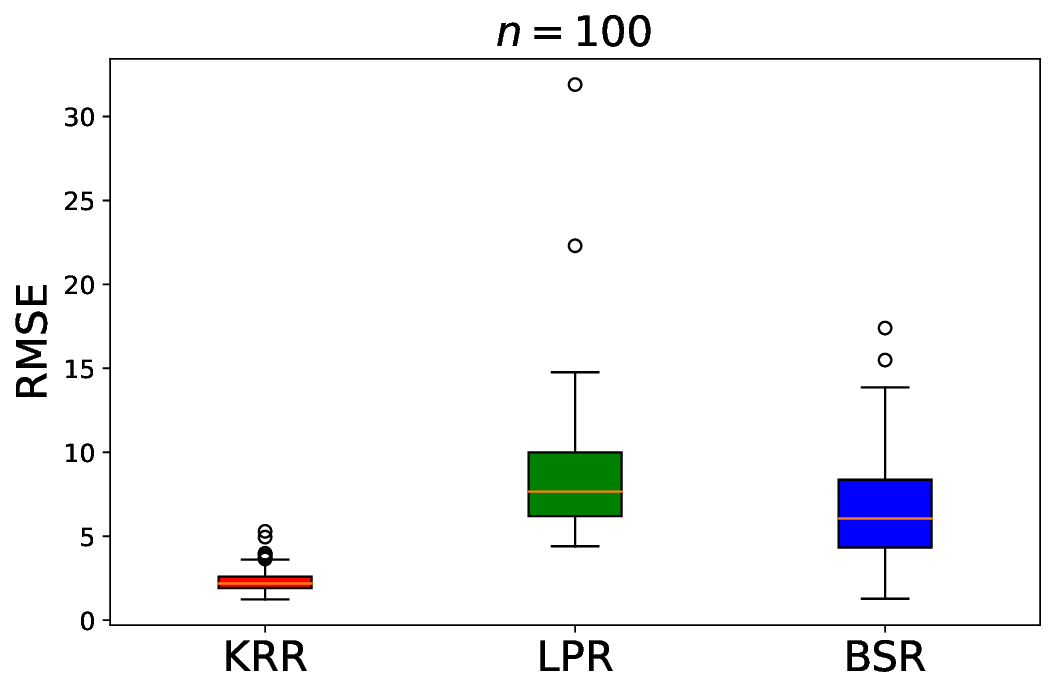}
  \includegraphics[width=1.97 in]{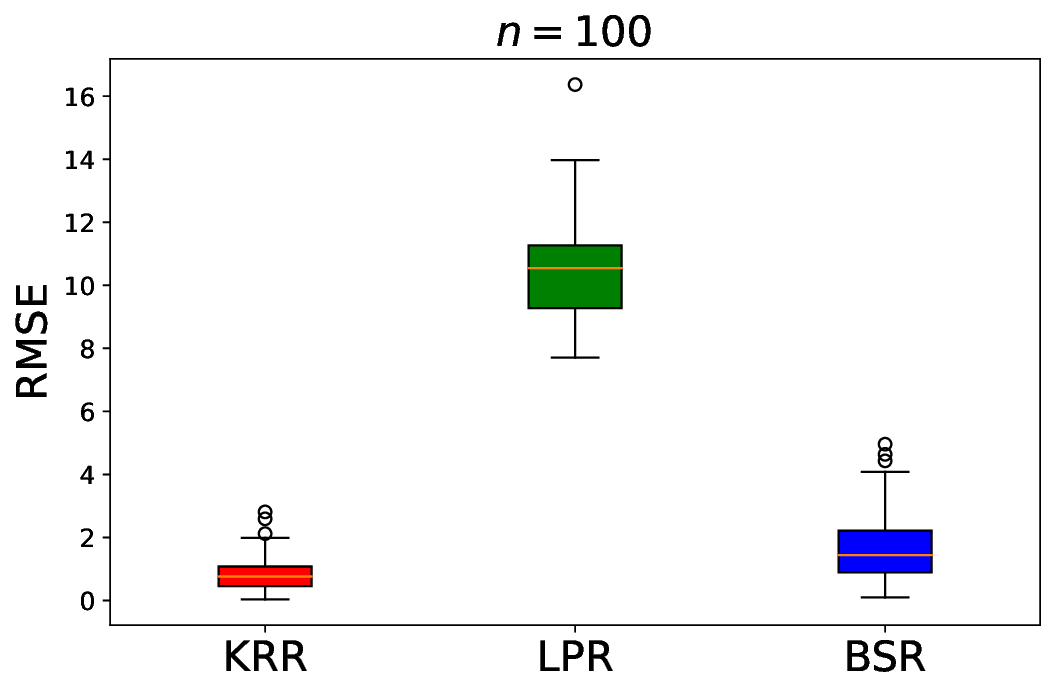}
   \includegraphics[width=1.97 in]{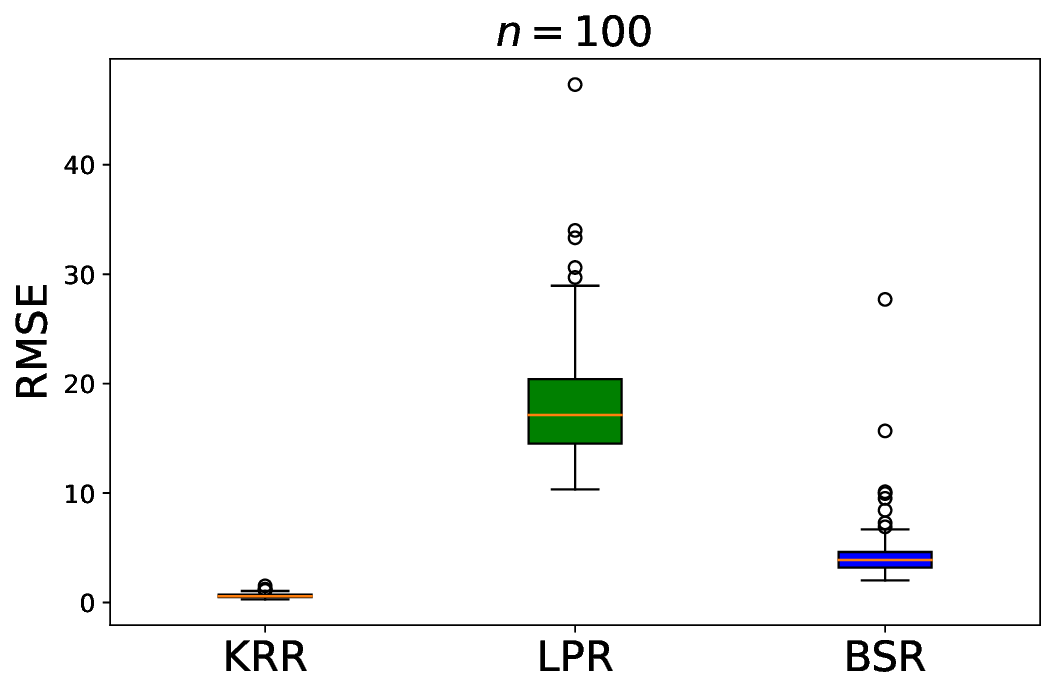}

   \includegraphics[width=1.97 in]{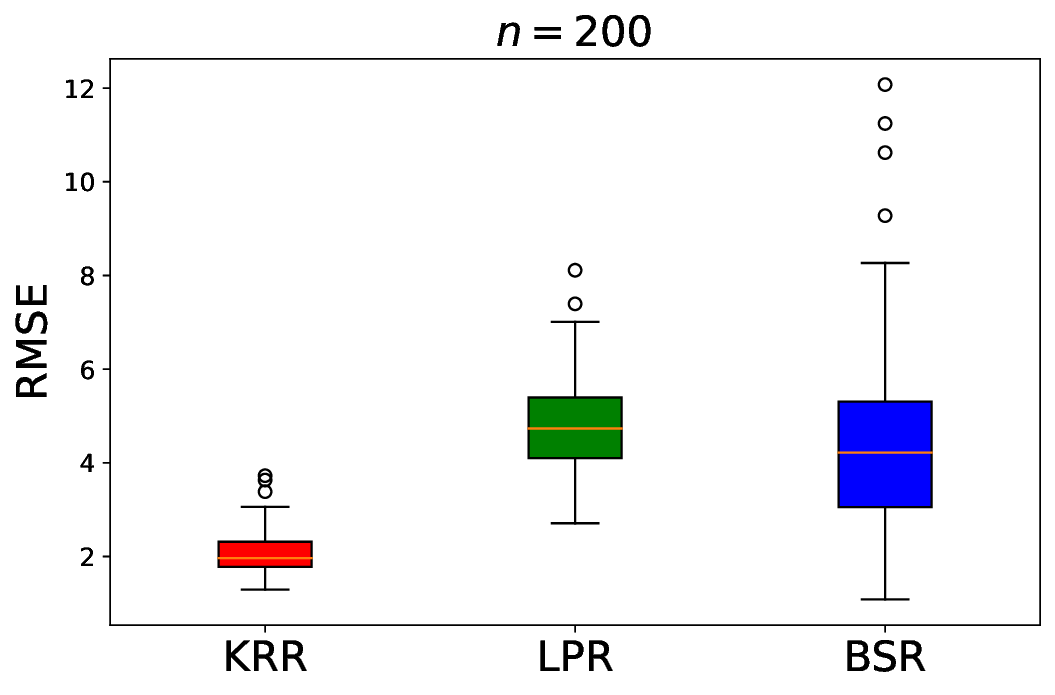}
  \includegraphics[width=1.97 in]{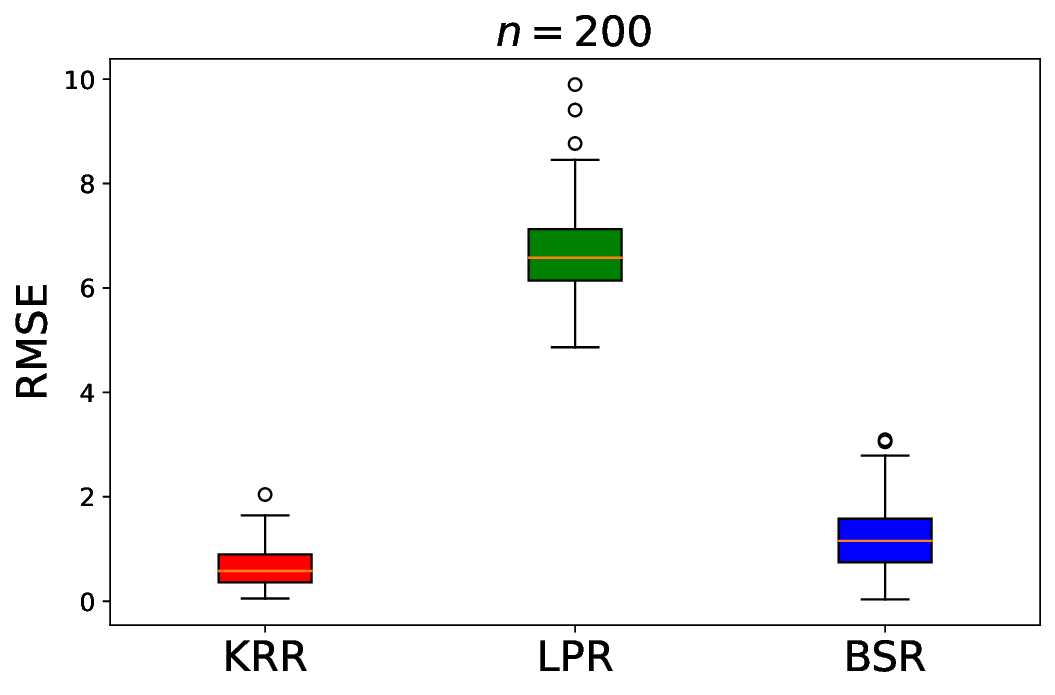}
   \includegraphics[width=1.97 in]{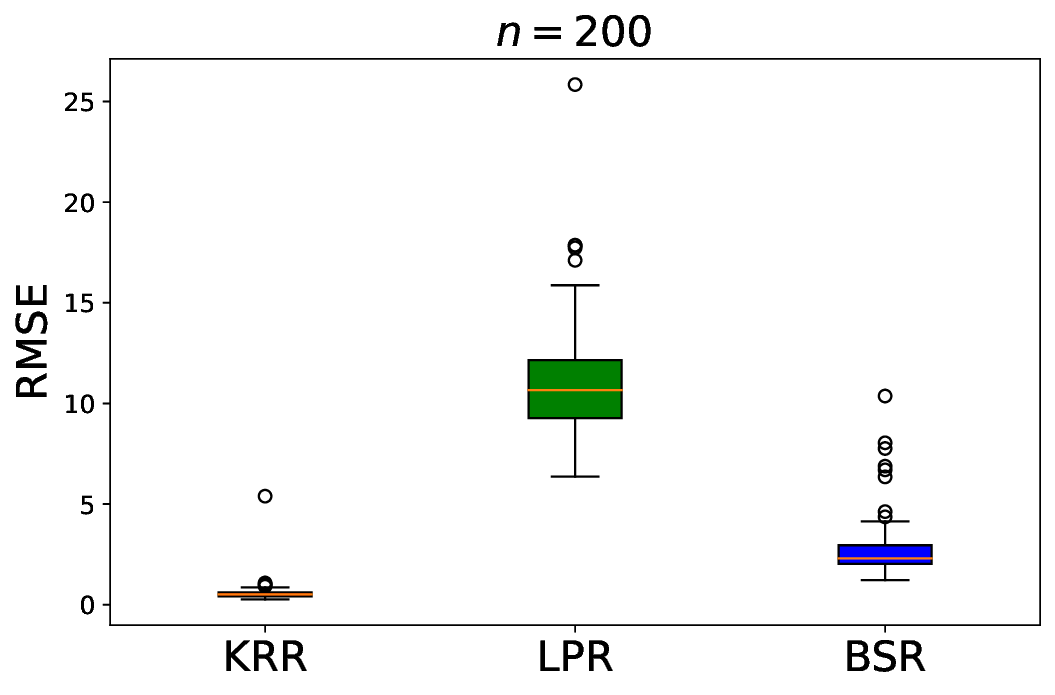}
       \caption{RMSE with $b=1$. Sample size: $n=100$ for top three panels and $n=200$ for bottom three panels.}
    \label{figure:test_sim}
\end{figure}

In this section, we evaluate the estimation performance of KRR estimator. For comparison, we also use local polynomial regression (LPR) and B-spline regression (BSR) estimators as the competitors. The bandwidth of LPR is selected by cross-validation. For BSR, we first construct cubic tensor product B-spline basis for DGP 1 and DGP 3, and additive B-spline basis for DGP 2. For each dimension, the number of knots is $6$, and the knots are equally spaced on $[0, 1]$. Next, we use ridge regression to fit the models, and the tuning parameter is selected by cross-validation. In each DGP, we let $b=1$, $\sigma=1$, and $n=100, 200$. The estimation accuracy is evaluated by the root mean squared errors (RMSE) defined as
\[
\textrm{RMSE} = \sqrt{\frac{1}{500}\sum_{i=1}^{500}\left(\partial^{\bfbeta} \widehat{f}(\bm{s}_i)-\partial^{\bfbeta}f_0(\bm{s}_i)\right)^2},
\]
where $\bm{s}_i, i=1, \ldots, 500$ are randomly generated on the domain of the covariates. The simulation results are summarized in Figure \ref{figure:test_sim}, which shows that the RMSEs decrease as $n$ increases.  Moreover, the KRR estimator has the best performance among three estimators.

\subsection{Bootstrap Hypothesis Testing}
\begin{figure}[ht!]
  \centering
  \includegraphics[width=1.97 in]{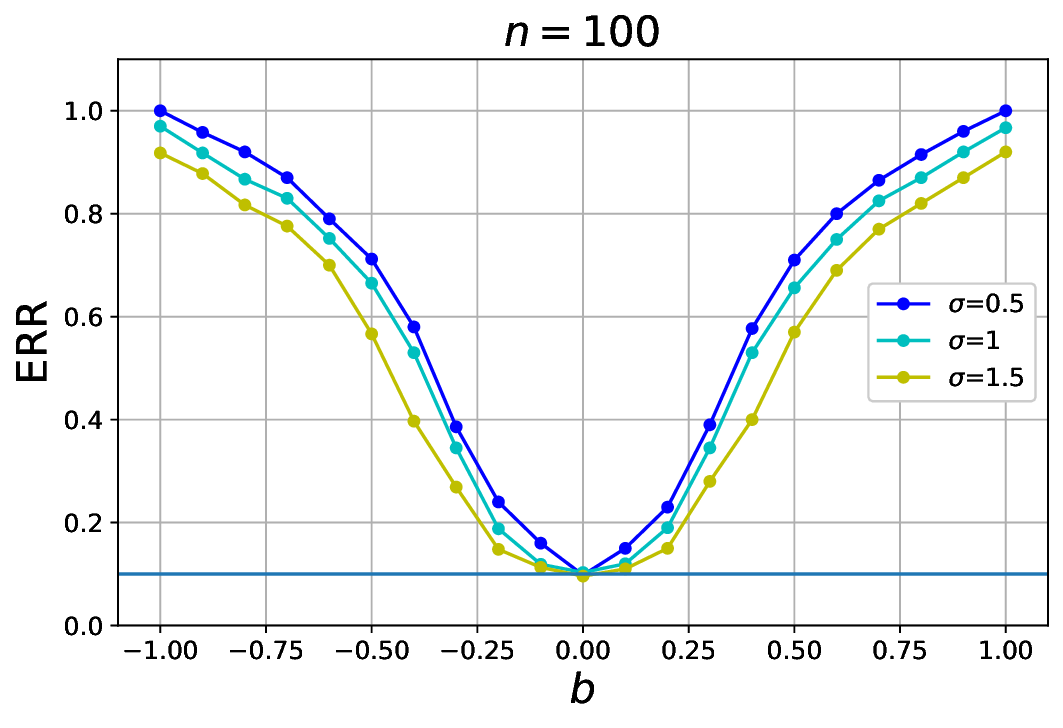}
  \includegraphics[width=1.97 in]{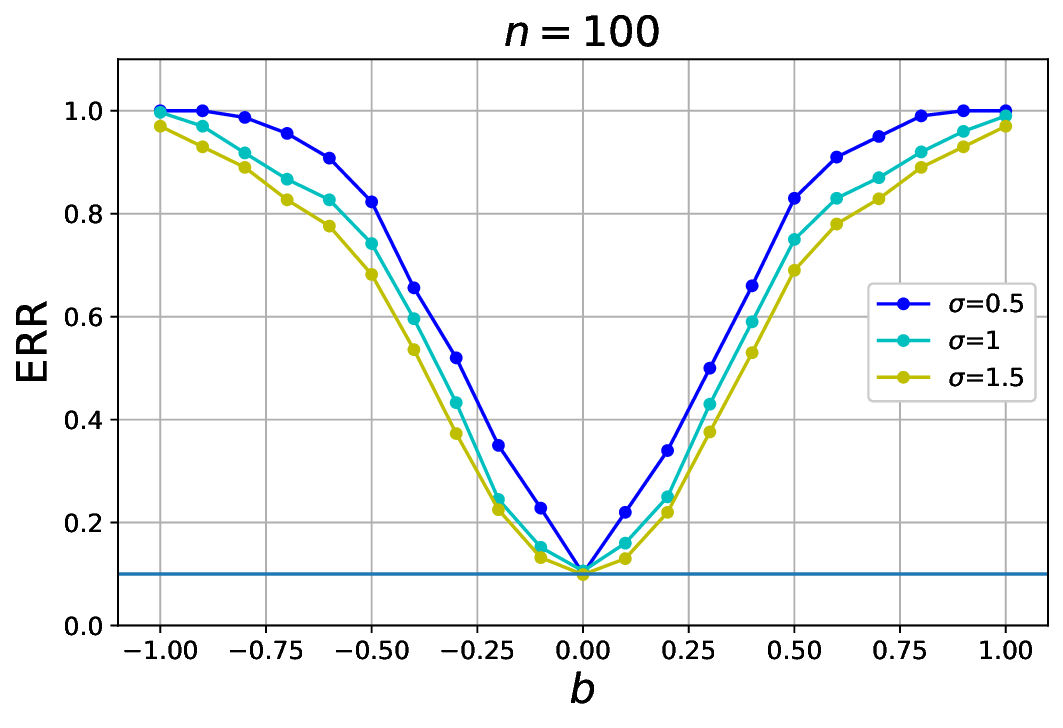}
  \includegraphics[width=1.97 in]{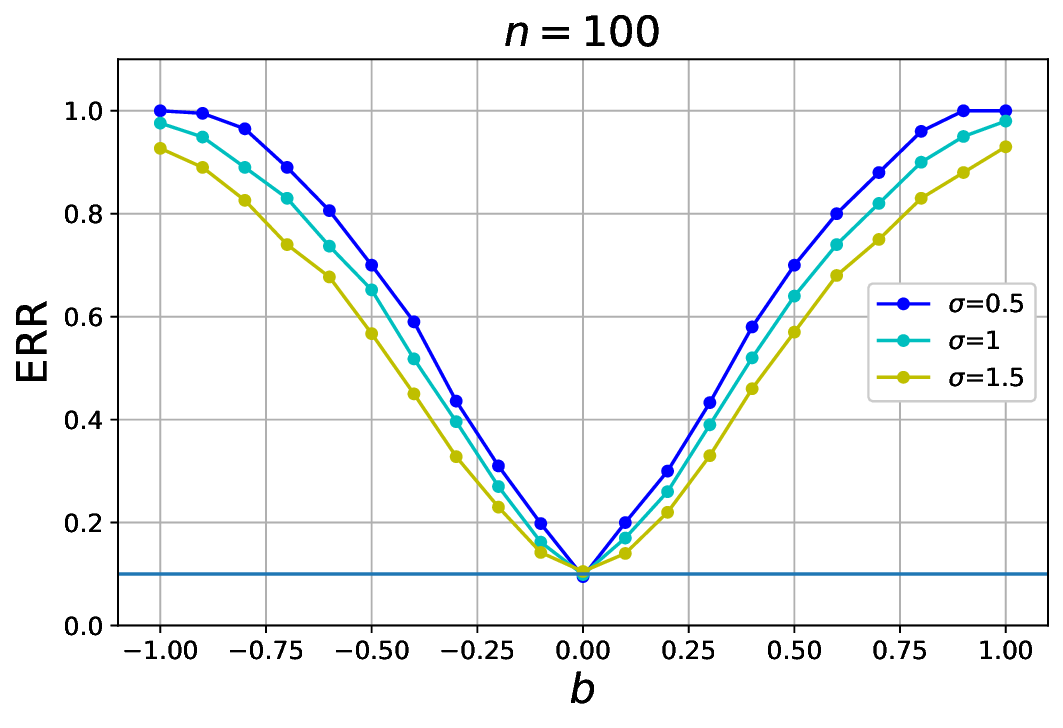}
  
  \includegraphics[width=1.97 in]{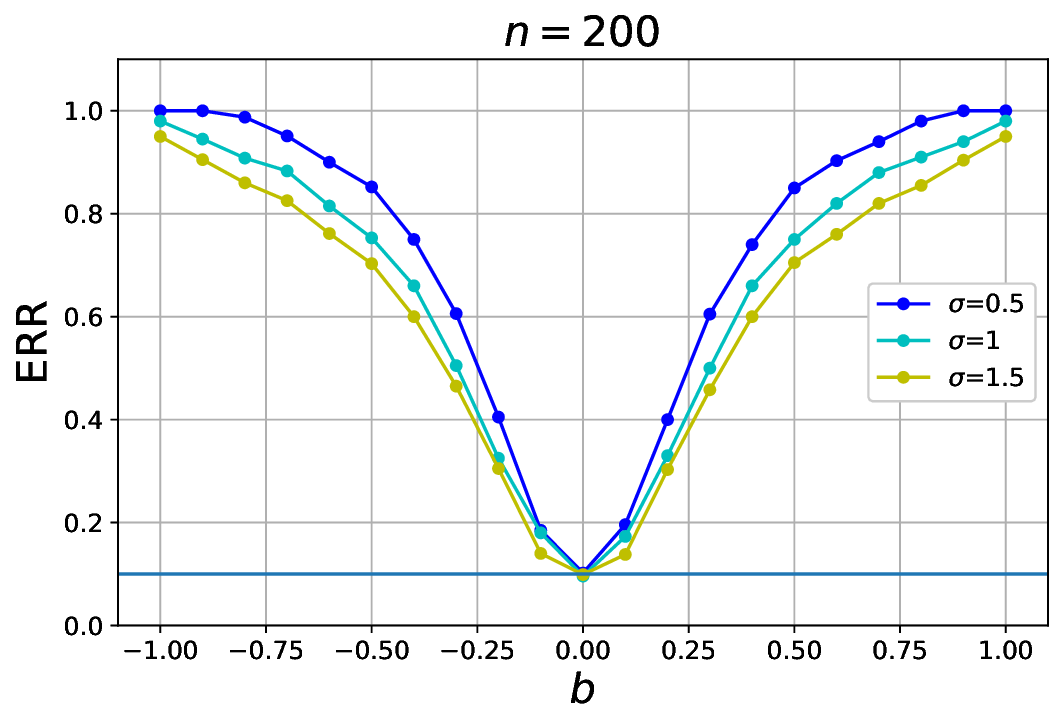}
   \includegraphics[width=1.97 in]{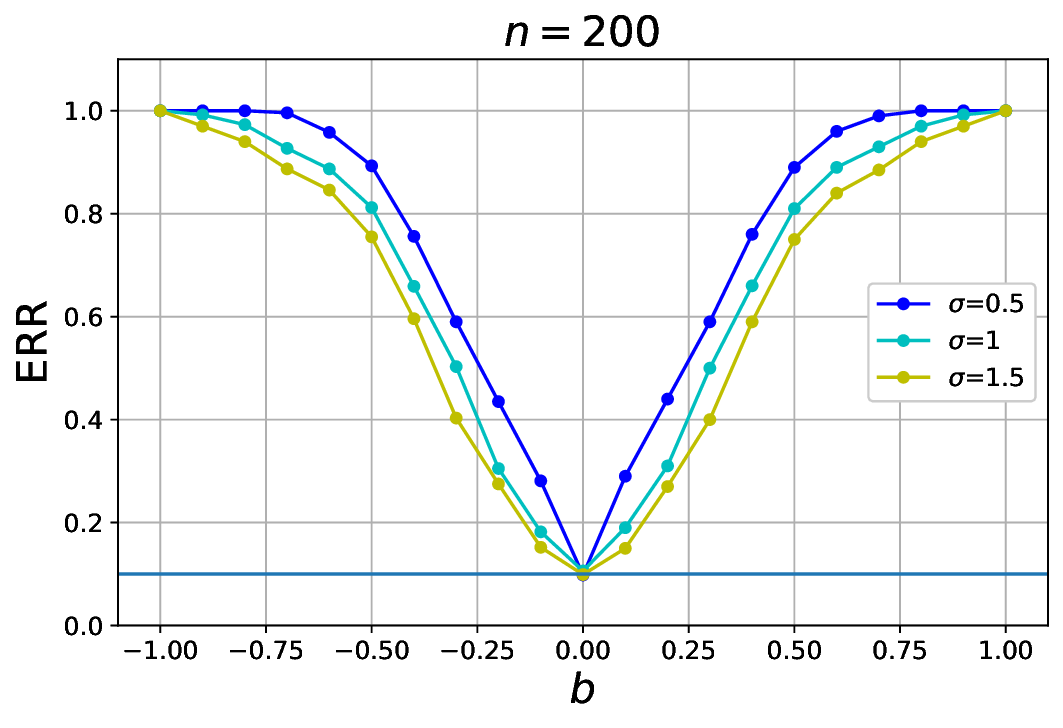}
    \includegraphics[width=1.97 in]{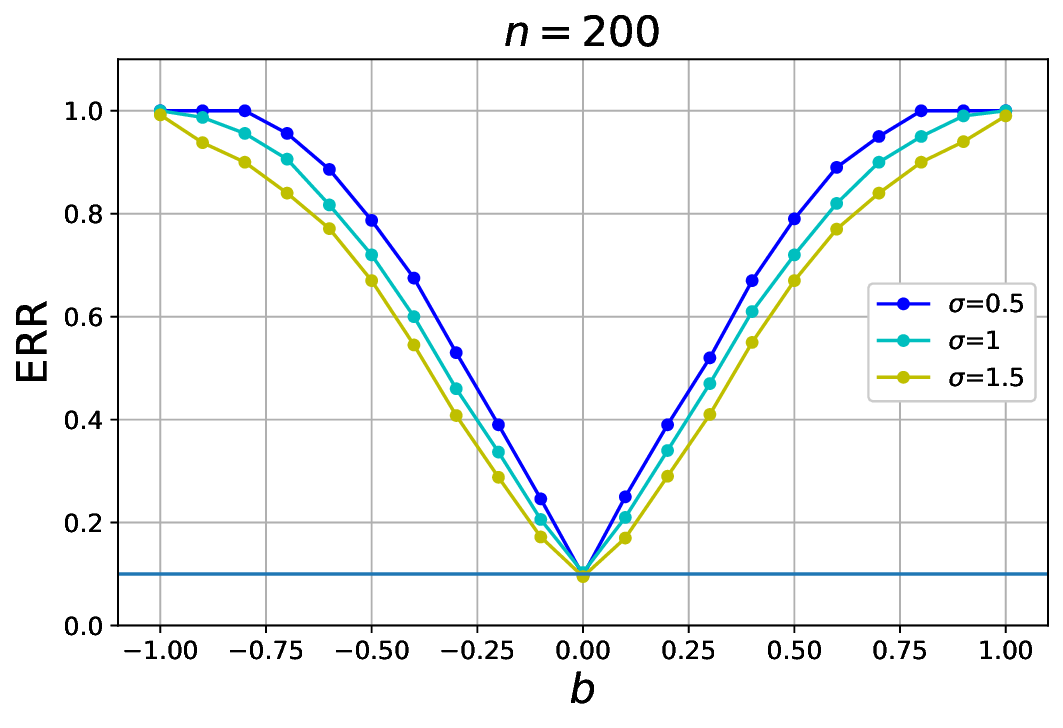}
   \caption{\it Empirical rejection rates with $\alpha=0.1$. DGP 1: left two panels, DGP 2: middle two panels, DGP 3: right two panels.}
    \label{figure:test_sim:dgp}
\end{figure}

In this section, we investigate the empirical sizes and powers of the bootstrap hypothesis testing procedure proposed in Algorithm \ref{alg:bootstrap}. Figures \ref{figure:test_sim:dgp} reports the empirical rejection rates (ERR) for each DGP with $\alpha=1$, $\sigma=0.5, 0.1, 1.5$, and $b\in [-1,1]$. It is straightforward to see that, when $b=0$, the rates are close to the nominal size, indicating the Type $\textnormal{\textup{I}}$  error could be well controlled as indicating be Theorem \ref{theorem:bootstrap:asymptotic}. When considering any value of $|b|$ greater than $0$, it can be observed that the EERs increase as the sample size $n$ grows larger. Given a sample size $n$, an increase in $|b|$ leads to an increase in the ERR. It is also easy to observe that as the signal-to-noise ratio increases (i.e., $\sigma$ decreases), the EER increases regardless of the sample size and $b$.

\section{Empirical Application}\label{section:empirical:application}

In this section, we apply the proposed method to the QSAR fish toxicity dataset (\citealp{cassotti2015similarity2})  from the UCI Machine Learning Repository. The dataset consists of information on 908 chemicals (n = 908). The response $(y)$ is the LC50 value, which represents the concentration of a chemical that causes the death of 50\% of the test fish over a test duration of 96 hours.  There are six molecular descriptors (predictors), including molecular properties ($X_1$), information indices ($X_2$), 2D autocorrelations ($X_3$), atom-type counts ($X_4$), atom-type counts ($X_5$), and 2D matrix-based descriptors ($X_6$). In our analysis, all the independent variables are standardized through dividing them by their maximum values. This standardization process ensures that all variables have a comparable scale and allows for a fair comparison and interpretation of their effects on the response variable. We perform a regression with six dependent variables $(r=6)$, and the highest order of interactions is set to six  $(q=6)$.  A series of hypothesis testing is conduced to identify the variables that exhibit a relationship with the response  and to explore the presence of any interactions between these variables. The bootstrap p-values (for testing derivatives with first- and second-order) are reported in Table \ref{tab:pvalues}.


\begin{table}[htbp]
  \centering
 \caption{P-values for hypothesis testing}
    \begin{tabular}{ccrrrrrrr}
    \hline
          & \multirow{2}[4]{*}{$H_0: \frac{\partial f_0}{\partial X_i}=0$} &       & \multicolumn{6}{c}{$H_0: \frac{\partial^2 f_0}{\partial X_i \partial X_j}=0$} \bigstrut\\
\cline{4-9}          &       &       & \multicolumn{1}{c}{$X_1$} & \multicolumn{1}{c}{$X_2$} & \multicolumn{1}{c}{$X_3$} & \multicolumn{1}{c}{$X_4$} & \multicolumn{1}{c}{$X_5$} & \multicolumn{1}{c}{$X_6$} \bigstrut\\
    \hline
    $X_1$   & \cellcolor[rgb]{ .867,  .922,  .969}0.006 &      & \;\;\;\;\;\;\;\;\;\cellcolor[rgb]{ .816,  .808,  .808} & 0.160      &  0.192     &  0.242     &   0.254    &  \multicolumn{1}{c}{\cellcolor[rgb]{ .867,  .922,  .969}0.086} \bigstrut[t]\\
    $X_2$    & \cellcolor[rgb]{ .867,  .922,  .969}0.010 &       & \cellcolor[rgb]{ .816,  .808,  .808} & \cellcolor[rgb]{ .816,  .808,  .808} & \multicolumn{1}{c}{\cellcolor[rgb]{ .867,  .922,  .969}0.082} &    0.204   &  0.238     & \multicolumn{1}{c}{\cellcolor[rgb]{ .867,  .922,  .969}0.074} \\
    $X_3$    & \cellcolor[rgb]{ .867,  .922,  .969}0.008 &       & \cellcolor[rgb]{ .816,  .808,  .808} & \cellcolor[rgb]{ .816,  .808,  .808} & \cellcolor[rgb]{ .816,  .808,  .808} &  0.270     &    0.232   & \multicolumn{1}{c}{\cellcolor[rgb]{ .867,  .922,  .969}0.066} \\
    $X_4$    & \cellcolor[rgb]{ .867,  .922,  .969}0.066 &       & \cellcolor[rgb]{ .816,  .808,  .808} & \cellcolor[rgb]{ .816,  .808,  .808} & \cellcolor[rgb]{ .816,  .808,  .808} & \cellcolor[rgb]{ .816,  .808,  .808} &   0.158    &  0.112\\
    $X_5$    &  0.144     &       & \cellcolor[rgb]{ .816,  .808,  .808} & \cellcolor[rgb]{ .816,  .808,  .808} & \cellcolor[rgb]{ .816,  .808,  .808} & \cellcolor[rgb]{ .816,  .808,  .808} & \cellcolor[rgb]{ .816,  .808,  .808} &  0.296\\
    $X_6$    & \cellcolor[rgb]{ .867,  .922,  .969}0.004 &       & \cellcolor[rgb]{ .816,  .808,  .808} & \cellcolor[rgb]{ .816,  .808,  .808} & \cellcolor[rgb]{ .816,  .808,  .808} & \cellcolor[rgb]{ .816,  .808,  .808} & \cellcolor[rgb]{ .816,  .808,  .808} & \cellcolor[rgb]{ .816,  .808,  .808} \bigstrut[b]\\
    \hline
    \end{tabular}%
  \label{tab:pvalues}%
\end{table}%

\begin{figure}[ht!]
  \centering
  \includegraphics[width=6 in]{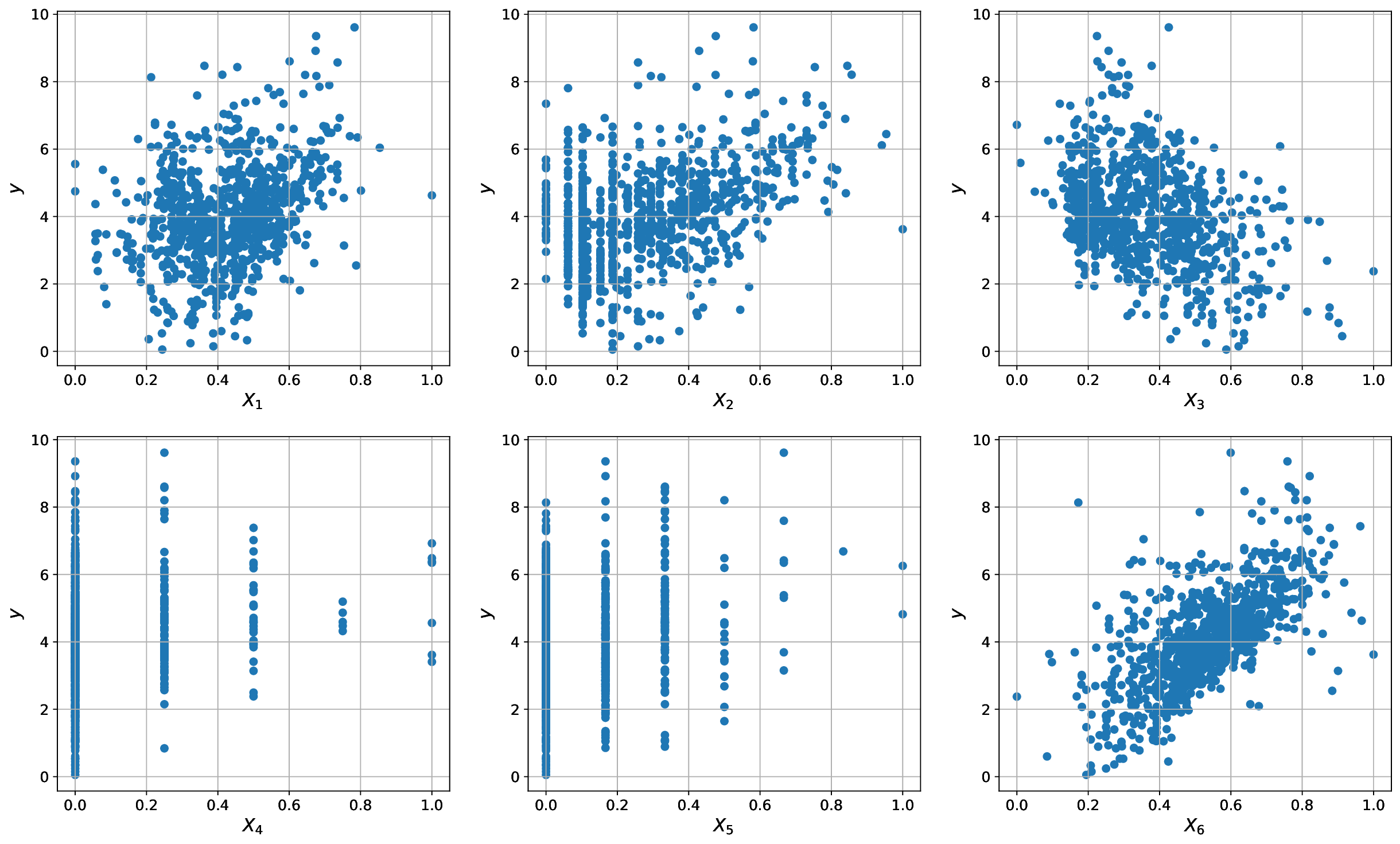}
   \caption{Scatter plots for response against predictors}
  \label{figure:real_scatter}
\end{figure}

Table \ref{tab:pvalues} reveals that $X_1, X_2, X_3, X_4, X_6$ are identified as the informative variables with  significance level  $\alpha=0.1$. This finding also can be confirmed by the  scatter plots in Figure \ref{figure:real_scatter}. It is evident that all the selected variables have a noticeable association with the response variable, while the correlation between $X_5$ and $y$ looks weaker.

\begin{figure}[ht!]
  \centering
  \includegraphics[width=6.2 in]{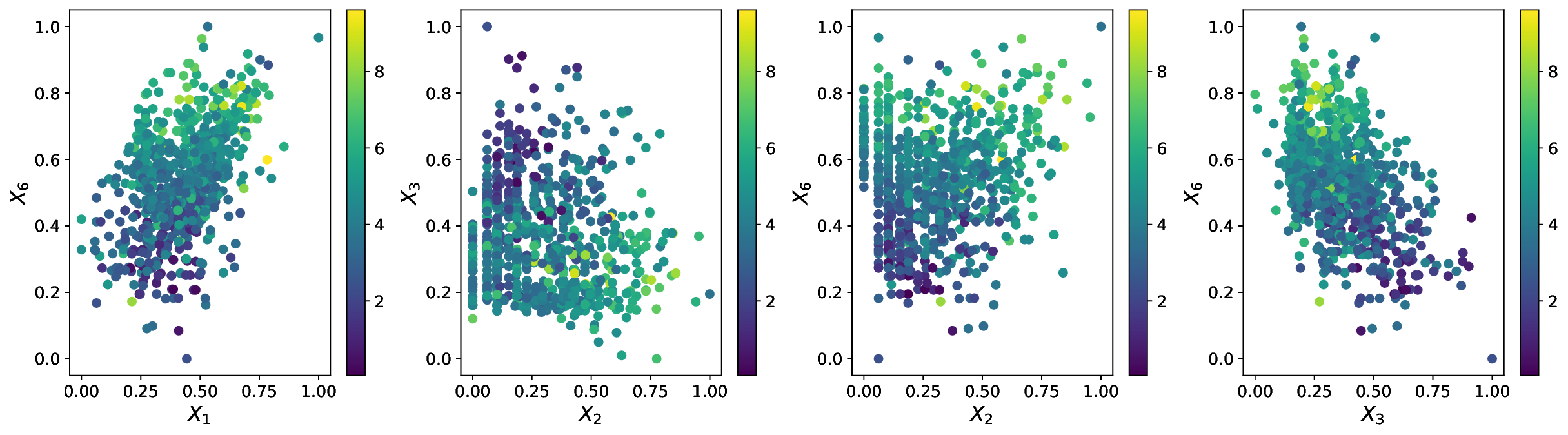}
   \caption{Heat maps for predictor pairs}
  \label{figure:real_interaction}
\end{figure}

Furthermore, Table  \ref{tab:pvalues} also suggests the presence of significant interactions for $(X_1, X_6)$, $(X_2, X_3)$, $(X_2, X_6)$, and $(X_3, X_6)$. Similar findings can be observed in the heat maps in Figure \ref{figure:real_interaction}.  For example, in the first subplot of Figure \ref{figure:real_interaction}, when both $X_1$ and $X_6$ are close to 1 (top right), the response shows a noticeably larger value compared to that when both $X_1$ and $X_6$ are small (bottom left). This observation suggests that there is a synergistic effect between $X_1$ and $X_6$ in influencing $y$.

We also examine the presence of high-order interactions (greater than or equal to three) between the covariates. After careful analysis, we discover that only the covariates $X_2, X_3, X_6$ exhibit a three-way interaction. The details are summarized in Table \ref{tab:additional:pvalue}.

\begin{table}[htbp]
  \centering
  \caption{P-values for testing of high-order derivatives}
  \resizebox{\columnwidth}{!}{%
    \begin{tabular}{lcccccccccccccccccccc}
    \hline
    Index & 123   & 124   & 125   & 126   & 134   & 135   & 136   & 145   & 146   & 146   & 234   & 245   & {\cellcolor[rgb]{ .867,  .922,  .969}236}  & 245   & 246   & 256   & 345   & 346   & 356   & 456 \bigstrut[t]\\
    P-value & 0.18  & 0.23  & 0.22  & 0.15  & 0.26  & 0.29  & 0.14  & 0.23  & 0.19  & 0.21  & 0.15  & 0.18  & {\cellcolor[rgb]{ .867,  .922,  .969}0.08}  & 0.25  & 0.27  & 0.17  & 0.22  & 0.15  & 0.16  & 0.2 \\
          &       &       &       &       &       &       &       &       &       &       &       &       &       &       &       &       &       &       &       &  \\
    Index & 1234  & 1235  & 1236  & 1245  & 1246  & 1256  & 1345  & 1346  & 1356  & 1456  & 2345  & 2346  & 2356  & 2456  & 3456  &       &       &       &       &  \\
    P-value & 0.39  & 0.34  & 0.39  & 0.4   & 0.47  & 0.32  & 0.42  & 0.43  & 0.4   & 0.37  & 0.44  & 0.37  & 0.43  & 0.35  & 0.42  &       &       &       &       &  \\
          &       &       &       &       &       &       &       &       &       &       &       &       &       &       &       &       &       &       &       &  \\
    Index & 12345 & 12346 & 12356 & 12456 & 13456 & 23456 & 123456 &       &       &       &       &       &       &       &       &       &       &       &       &  \\
    P-value & 0.48  & 0.44  & 0.45  & 0.52  & 0.43  & 0.44  & 0.52  &       &       &       &       &       &       &       &       &       &       &       &       &  \bigstrut[b]\\
    \hline
    \end{tabular}%
  \label{tab:additional:pvalue}%
}  \par
{\vspace{0.5em} {\footnotesize Note: Index $i_1\ldots i_k$ corresponds to testing null hypothesis $H_0: \frac{\partial^k f_0}{\partial X_{i_1} \ldots \partial X_{i_k}}=0$.}} 
\end{table}%

In the final evaluation, we compare the prediction errors of three models whose specifications are summarized in Table \ref{tab:prediciton:error}. The entire dataset is randomly divided by a $60-40$ split for training and prediction, and the averaged prediction errors and the standard deviations are based on 100 replications. Table \ref{tab:prediciton:error} shows that the additive model has the worst performance, while the reduced model performs similarly to the full model, which validates the previous analysis of hypothesis testing.

\begin{table}[ht!]
  \centering
  \caption{Prediction errors}
    \begin{tabular}{lccc}
    \hline
          & Variables & Interaction Order $(q)$ & Error \bigstrut[t]\\
          \hline 
    Full Model &    $(X_1,X_2, X_3, X_4, X_5, X_6)$   & 6     &  0.273 (0.07) \\
    Additive Model &    $(X_1,X_2, X_3, X_4, X_5, X_6)$   & 1     & 1.21 (0.27) \\
    Reduced Model &    $(X_1,X_2, X_3, X_4, X_6)$   & 3     & 0.282 (0.05) \bigstrut[b]\\
    \hline
    \end{tabular}%
  \label{tab:prediciton:error}%
\end{table}%

\section{Conclusion}
In this paper, we propose a plug-in KRR estimator for estimating derivatives of the underlying regression function in SS-ANOVA models. The proposed estimator can be easily calculated and enjoys favorable theoretical properties. We first establish an $L_\infty$ convergence rate of the proposed estimator under general random designs. Additionally, when the covariates are uniformly distributed, we obtain a sharp $L_2$ convergence rate under certain order conditions on the derivatives. Motivated by the wide range of real-world applications, we introduce a hypothesis testing procedure to examine whether a partial derivative is zero. To address the difficulty in estimating the critical value of the test, we also develop an associated bootstrap algorithm to construct the rejection region and calculate the p-value.

There are several directions for future work. For example, a natural step is to derive the weak convergence of the process $\bfx\to \partial^{\bfbeta}\widehat{f}(\bfx)$, which allows to construct simultaneous confidence corridors. It is also of interest to extend the minimax lower bound in Theorem \ref{theorem:lower:bound:equal:beta} to general $\bfbeta$. Finally, our analysis suggests a path toward model-free variable selection methods for interpretable learning (see, e.g., \citealp{RODEO, knockoff, YeLi2025, YeSenftleLi2024}), where the goal is to reliably identify a small set of relevant covariates in flexible, nonparametric settings.

\bibliographystyle{Chicago}
\bibliography{ref}

\begin{thebibliography}{}

\bibitem[\protect\citeauthoryear{An, Liu, and Venkatesh}{An
  et~al.}{2007}]{an2007face}
An, S., W.~Liu, and S.~Venkatesh (2007).
\newblock Face recognition using kernel ridge regression.
\newblock In {\em 2007 IEEE Conference on Computer Vision and Pattern
  Recognition}, pp.\  1--7. IEEE.

\bibitem[\protect\citeauthoryear{Atalla, Gasim, and Hunt}{Atalla
  et~al.}{2018}]{atalla2018gasoline}
Atalla, T.~N., A.~A. Gasim, and L.~C. Hunt (2018).
\newblock Gasoline demand, pricing policy, and social welfare in saudi arabia:
  A quantitative analysis.
\newblock {\em Energy policy\/}~{\em 114}, 123--133.

\bibitem[\protect\citeauthoryear{Ballabio, Cassotti, Consonni, and
  Todeschini}{Ballabio et~al.}{2019}]{cassotti2015similarity2}
Ballabio, D., M.~Cassotti, V.~Consonni, and R.~Todeschini (2019).
\newblock Qsar fish toxicity.
\newblock UCI Machine Learning Repository.
\newblock {DOI}: https://doi.org/10.24432/C5JG7B.

\bibitem[\protect\citeauthoryear{Banerjee, Gelfand, and Sirmans}{Banerjee
  et~al.}{2003}]{banerjee2003directional}
Banerjee, S., A.~E. Gelfand, and C.~Sirmans (2003).
\newblock Directional rates of change under spatial process models.
\newblock {\em Journal of the American Statistical Association\/}~{\em
  98\/}(464), 946--954.

\bibitem[\protect\citeauthoryear{Blundell, Duncan, and Pendakur}{Blundell
  et~al.}{1998}]{blundell1998semiparametric}
Blundell, R., A.~Duncan, and K.~Pendakur (1998).
\newblock Semiparametric estimation and consumer demand.
\newblock {\em Journal of Applied econometrics\/}~{\em 13\/}(5), 435--461.

\bibitem[\protect\citeauthoryear{Cai and Yuan}{Cai and
  Yuan}{2012}]{cai2012minimax}
Cai, T.~T. and M.~Yuan (2012).
\newblock Minimax and adaptive prediction for functional linear regression.
\newblock {\em Journal of the American Statistical Association\/}~{\em
  107\/}(499), 1201--1216.

\bibitem[\protect\citeauthoryear{Calonico, Cattaneo, and Farrell}{Calonico
  et~al.}{2018}]{calonico2018effect}
Calonico, S., M.~D. Cattaneo, and M.~H. Farrell (2018).
\newblock On the effect of bias estimation on coverage accuracy in
  nonparametric inference.
\newblock {\em Journal of the American Statistical Association\/}~{\em
  113\/}(522), 767--779.

\bibitem[\protect\citeauthoryear{Cand\`{e}s, Fan, Janson, and Lv}{Cand\`{e}s
  et~al.}{2018}]{knockoff}
Cand\`{e}s, E., Y.~Fan, L.~Janson, and J.~Lv (2018, 01).
\newblock {Panning for Gold: ‘Model-X’ Knockoffs for High Dimensional
  Controlled Variable Selection}.
\newblock {\em Journal of the Royal Statistical Society Series B: Statistical
  Methodology\/}~{\em 80\/}(3), 551--577.

\bibitem[\protect\citeauthoryear{Charnes, Cooper, Golany, Seiford, and
  Stutz}{Charnes et~al.}{1985}]{charnes1985foundations}
Charnes, A., W.~W. Cooper, B.~Golany, L.~Seiford, and J.~Stutz (1985).
\newblock Foundations of data envelopment analysis for pareto-koopmans
  efficient empirical production functions.
\newblock {\em Journal of Econometrics\/}~{\em 30\/}(1-2), 91--107.

\bibitem[\protect\citeauthoryear{Chen, McIlroy, Archana, Baker, and
  Panagiotou}{Chen et~al.}{2019}]{chen2019pollution}
Chen, J., S.~E. McIlroy, A.~Archana, D.~M. Baker, and G.~Panagiotou (2019).
\newblock A pollution gradient contributes to the taxonomic, functional, and
  resistome diversity of microbial communities in marine sediments.
\newblock {\em Microbiome\/}~{\em 7}, 1--12.

\bibitem[\protect\citeauthoryear{Cheng and Shang}{Cheng and Shang}{2015}]{cs15}
Cheng, G. and Z.~Shang (2015).
\newblock Joint asymptotics for semi-nonparametric regression models with
  partially linear structure.
\newblock {\em Annals of Statistics\/}~{\em 43\/}(3), 1351--1390.

\bibitem[\protect\citeauthoryear{Chernozhukov, Chetverikov, and
  Kato}{Chernozhukov et~al.}{2016}]{chernozhukov2016empirical}
Chernozhukov, V., D.~Chetverikov, and K.~Kato (2016).
\newblock Empirical and multiplier bootstraps for suprema of empirical
  processes of increasing complexity, and related gaussian couplings.
\newblock {\em Stochastic Processes and their Applications\/}~{\em 126\/}(12),
  3632--3651.

\bibitem[\protect\citeauthoryear{Cristianini and Shawe-Taylor}{Cristianini and
  Shawe-Taylor}{2000}]{cristianini2000introduction}
Cristianini, N. and J.~Shawe-Taylor (2000).
\newblock {\em An Introduction to Support Vector Machines and Other
  Kernel-based Learning Methods}.
\newblock Cambridge university press.

\bibitem[\protect\citeauthoryear{Dai and Chien}{Dai and
  Chien}{2017}]{dai2017minimax}
Dai, X. and P.~Chien (2017).
\newblock Minimax optimal rates of estimation in functional anova models with
  derivatives.
\newblock {\em arXiv preprint arXiv:1706.00850\/}.

\bibitem[\protect\citeauthoryear{Deaton}{Deaton}{1986}]{deaton1986demand}
Deaton, A. (1986).
\newblock Demand analysis.
\newblock Volume~3 of {\em Handbook of Econometrics}, pp.\  1767--1839.
  Elsevier.

\bibitem[\protect\citeauthoryear{Dong, Chen, and Pan}{Dong
  et~al.}{2017}]{dong2017learning}
Dong, X., S.~Chen, and S.~Pan (2017).
\newblock Learning to prune deep neural networks via layer-wise optimal brain
  surgeon.
\newblock {\em Advances in Neural Information Processing Systems\/}~{\em 30}.

\bibitem[\protect\citeauthoryear{Duhr and Braun}{Duhr and
  Braun}{2006}]{duhr2006molecules}
Duhr, S. and D.~Braun (2006).
\newblock Why molecules move along a temperature gradient.
\newblock {\em Proceedings of the National Academy of Sciences\/}~{\em
  103\/}(52), 19678--19682.

\bibitem[\protect\citeauthoryear{Exterkate, Groenen, Heij, and van
  Dijk}{Exterkate et~al.}{2016}]{exterkate2016nonlinear}
Exterkate, P., P.~J. Groenen, C.~Heij, and D.~van Dijk (2016).
\newblock Nonlinear forecasting with many predictors using kernel ridge
  regression.
\newblock {\em International Journal of Forecasting\/}~{\em 32\/}(3), 736--753.

\bibitem[\protect\citeauthoryear{Gao, Wahba, Klein, and Klein}{Gao
  et~al.}{2001}]{gao2001smoothing}
Gao, F., G.~Wahba, R.~Klein, and B.~Klein (2001).
\newblock Smoothing spline anova for multivariate bernoulli observations with
  application to ophthalmology data.
\newblock {\em Journal of the American Statistical Association\/}~{\em
  96\/}(453), 127--160.

\bibitem[\protect\citeauthoryear{Gorsich and Genton}{Gorsich and
  Genton}{2000}]{gorsich2000variogram}
Gorsich, D.~J. and M.~G. Genton (2000).
\newblock Variogram model selection via nonparametric derivative estimation.
\newblock {\em Mathematical geology\/}~{\em 32}, 249--270.

\bibitem[\protect\citeauthoryear{Gu}{Gu}{2013}]{gu2013smoothing}
Gu, C. (2013).
\newblock {\em Smoothing Spline ANOVA Models}.
\newblock Springer New York, NY.

\bibitem[\protect\citeauthoryear{Hassibi and Stork}{Hassibi and
  Stork}{1992}]{hassibi1992second}
Hassibi, B. and D.~Stork (1992).
\newblock Second order derivatives for network pruning: Optimal brain surgeon.
\newblock {\em Advances in neural information processing systems\/}~{\em 5}.

\bibitem[\protect\citeauthoryear{Hastie, Tibshirani, and Friedman}{Hastie
  et~al.}{2009}]{hastie2009elements}
Hastie, T., R.~Tibshirani, and J.~Friedman (2009).
\newblock {\em The Elements of Statistical Learning: Data mining, Inference,
  and Prediction}.
\newblock Springer Science \& Business Media.

\bibitem[\protect\citeauthoryear{Haworth, Shawe-Taylor, Cheng, and
  Wang}{Haworth et~al.}{2014}]{haworth2014local}
Haworth, J., J.~Shawe-Taylor, T.~Cheng, and J.~Wang (2014).
\newblock Local online kernel ridge regression for forecasting of urban travel
  times.
\newblock {\em Transportation research part C: emerging technologies\/}~{\em
  46}, 151--178.

\bibitem[\protect\citeauthoryear{Hu and Zastawniak}{Hu and
  Zastawniak}{2020}]{hu2020pricing}
Hu, W. and T.~Zastawniak (2020).
\newblock Pricing high-dimensional american options by kernel ridge regression.
\newblock {\em Quantitative Finance\/}~{\em 20\/}(5), 851--865.

\bibitem[\protect\citeauthoryear{Huang}{Huang}{1998}]{h98}
Huang, J. (1998).
\newblock Projection estimation in multiple regression with application to
  functional anova models.
\newblock {\em Annals of Statistics\/}~{\em 26\/}(1), 242--272.

\bibitem[\protect\citeauthoryear{Kumar and Aravind}{Kumar and
  Aravind}{2008}]{kumar2008face}
Kumar, B.~V. and R.~Aravind (2008).
\newblock Face hallucination using olpp and kernel ridge regression.
\newblock In {\em 2008 15th IEEE International Conference on Image Processing},
  pp.\  353--356. IEEE.

\bibitem[\protect\citeauthoryear{Labandeira, Labeaga, and
  L{\'o}pez-Otero}{Labandeira et~al.}{2017}]{labandeira2017meta}
Labandeira, X., J.~M. Labeaga, and X.~L{\'o}pez-Otero (2017).
\newblock A meta-analysis on the price elasticity of energy demand.
\newblock {\em Energy policy\/}~{\em 102}, 549--568.

\bibitem[\protect\citeauthoryear{Lafferty and Wasserman}{Lafferty and
  Wasserman}{2008}]{RODEO}
Lafferty, J. and L.~Wasserman (2008).
\newblock {Rodeo: Sparse, greedy nonparametric regression}.
\newblock {\em The Annals of Statistics\/}~{\em 36\/}(1), 28--63.

\bibitem[\protect\citeauthoryear{Lian}{Lian}{2022}]{lian2022distributed}
Lian, H. (2022).
\newblock Distributed learning of conditional quantiles in the reproducing
  kernel hilbert space.
\newblock {\em Advances in Neural Information Processing Systems\/}~{\em 35},
  11686--11696.

\bibitem[\protect\citeauthoryear{Lin}{Lin}{2000}]{lin2000tensor}
Lin, Y. (2000).
\newblock Tensor product space anova models.
\newblock {\em Annals of Statistics\/}~{\em 28\/}(3), 734--755.

\bibitem[\protect\citeauthoryear{Liu, Shang, and Cheng}{Liu
  et~al.}{2019}]{pmlr-v99-liu19a}
Liu, M., Z.~Shang, and G.~Cheng (2019).
\newblock Sharp theoretical analysis for nonparametric testing under random
  projection.
\newblock In {\em Conference on Learning Theory}, Volume~99, pp.\  2175--2209.

\bibitem[\protect\citeauthoryear{Liu, Shang, and Cheng}{Liu
  et~al.}{2020}]{liu2020distributed}
Liu, M., Z.~Shang, and G.~Cheng (2020).
\newblock {Nonparametric distributed learning under general designs}.
\newblock {\em Electronic Journal of Statistics\/}~{\em 14\/}(2), 3070 -- 3102.

\bibitem[\protect\citeauthoryear{Liu and De~Brabanter}{Liu and
  De~Brabanter}{2018}]{liu2018derivative}
Liu, Y. and K.~De~Brabanter (2018).
\newblock Derivative estimation in random design.
\newblock {\em Advances in Neural Information Processing Systems\/}~{\em 31}.

\bibitem[\protect\citeauthoryear{Liu and De~Brabanter}{Liu and
  De~Brabanter}{2020}]{liu2020smoothed}
Liu, Y. and K.~De~Brabanter (2020).
\newblock Smoothed nonparametric derivative estimation using weighted
  difference quotients.
\newblock {\em Journal of Machine Learning Research\/}~{\em 21\/}(1),
  2438--2482.

\bibitem[\protect\citeauthoryear{Liu and Li}{Liu and
  Li}{2023}]{liu2020estimation}
Liu, Z. and M.~Li (2023).
\newblock On the estimation of derivatives using plug-in kernel ridge
  regression estimators.
\newblock {\em Journal of Machine Learning Research\/}~{\em 24\/}(266), 1--37.

\bibitem[\protect\citeauthoryear{Liu and Li}{Liu and Li}{2026}]{liu2026optimal}
Liu, Z. and M.~Li (2026).
\newblock Optimal plug-in {G}aussian processes for modeling derivatives.
\newblock {\em Journal of the American Statistical Association\/}, 1--22.

\bibitem[\protect\citeauthoryear{Luo, Wahba, and Johnson}{Luo
  et~al.}{1998}]{luo1998spatial}
Luo, Z., G.~Wahba, and D.~R. Johnson (1998).
\newblock Spatial--temporal analysis of temperature using smoothing spline
  anova.
\newblock {\em Journal of Climate\/}~{\em 11\/}(1), 18--28.

\bibitem[\protect\citeauthoryear{Meron, Gilad, Von~Hardenberg, Shachak, and
  Zarmi}{Meron et~al.}{2004}]{meron2004vegetation}
Meron, E., E.~Gilad, J.~Von~Hardenberg, M.~Shachak, and Y.~Zarmi (2004).
\newblock Vegetation patterns along a rainfall gradient.
\newblock {\em Chaos, Solitons \& Fractals\/}~{\em 19\/}(2), 367--376.

\bibitem[\protect\citeauthoryear{Reich, Storlie, and Bondell}{Reich
  et~al.}{2009}]{reich2009variable}
Reich, B.~J., C.~B. Storlie, and H.~D. Bondell (2009).
\newblock Variable selection in bayesian smoothing spline anova models:
  Application to deterministic computer codes.
\newblock {\em Technometrics\/}~{\em 51\/}(2), 110--120.

\bibitem[\protect\citeauthoryear{Sang, Shang, and Du}{Sang
  et~al.}{2022}]{sang2022statistical}
Sang, P., Z.~Shang, and P.~Du (2022).
\newblock Statistical inference for functional linear quantile regression.
\newblock {\em arXiv preprint arXiv:2202.11747\/}.

\bibitem[\protect\citeauthoryear{Shang}{Shang}{2010}]{shangejs2010}
Shang, Z. (2010).
\newblock {Convergence rate and Bahadur type representation of general
  smoothing spline M-estimates}.
\newblock {\em Electronic Journal of Statistics\/}~{\em 4\/}(none), 1411 --
  1442.

\bibitem[\protect\citeauthoryear{Shang and Cheng}{Shang and
  Cheng}{2013}]{sc13aos}
Shang, Z. and G.~Cheng (2013).
\newblock Local and global asymptotic inference in smoothing spline models.
\newblock {\em Annals of Statistics\/}~{\em 41\/}(5), 2608--2638.

\bibitem[\protect\citeauthoryear{Shang and Cheng}{Shang and
  Cheng}{2015}]{shangaos2015}
Shang, Z. and G.~Cheng (2015).
\newblock {Nonparametric inference in generalized functional linear models}.
\newblock {\em Annals of Statistics\/}~{\em 43\/}(4), 1742 -- 1773.

\bibitem[\protect\citeauthoryear{Shephard}{Shephard}{2015}]{shephard2015theory}
Shephard, R.~W. (2015).
\newblock {\em Theory of Cost and Production Functions}.
\newblock Princeton University Press.

\bibitem[\protect\citeauthoryear{Steinwart, Hush, Scovel, et~al.}{Steinwart
  et~al.}{2009}]{steinwart2009optimal}
Steinwart, I., D.~R. Hush, C.~Scovel, et~al. (2009).
\newblock Optimal rates for regularized least squares regression.
\newblock In {\em Conference on Learning Theory}, pp.\  79--93.

\bibitem[\protect\citeauthoryear{Stone}{Stone}{1982}]{stone1982optimal}
Stone, C.~J. (1982).
\newblock Optimal global rates of convergence for nonparametric regression.
\newblock {\em Annals of Statistics\/}, 1040--1053.

\bibitem[\protect\citeauthoryear{Sun, Du, Wang, and Ma}{Sun
  et~al.}{2018}]{sun2018optimal}
Sun, X., P.~Du, X.~Wang, and P.~Ma (2018).
\newblock Optimal penalized function-on-function regression under a reproducing
  kernel hilbert space framework.
\newblock {\em Journal of the American Statistical Association\/}~{\em
  113\/}(524), 1601--1611.

\bibitem[\protect\citeauthoryear{Touzani and Busby}{Touzani and
  Busby}{2013}]{touzani2013smoothing}
Touzani, S. and D.~Busby (2013).
\newblock Smoothing spline analysis of variance approach for global sensitivity
  analysis of computer codes.
\newblock {\em Reliability Engineering \& System Safety\/}~{\em 112}, 67--81.

\bibitem[\protect\citeauthoryear{Tuo and Zou}{Tuo and
  Zou}{2024}]{TuoZou2024KRRFunctionals}
Tuo, R. and L.~Zou (2024).
\newblock Asymptotic theory for linear functionals of kernel ridge regression.
\newblock {\em Statistica Sinica\/}.
\newblock In press.

\bibitem[\protect\citeauthoryear{Wahba}{Wahba}{2003}]{WAHBA2003531}
Wahba, G. (2003).
\newblock An introduction to smoothing spline anova models in rkhs, with
  examples in geographical data, medicine, atmospheric sciences and machine
  learning.
\newblock {\em IFAC Proceedings Volumes\/}~{\em 36\/}(16), 531--536.
\newblock 13th IFAC Symposium on System Identification (SYSID 2003), Rotterdam,
  The Netherlands, 27-29 August, 2003.

\bibitem[\protect\citeauthoryear{Wahba and Luo}{Wahba and
  Luo}{1996}]{wahba1996smoothing}
Wahba, G. and Z.~Luo (1996).
\newblock Smoothing spline anova fits for very large, nearly regular data sets,
  with application to historical global climate data.
\newblock {\em Annals of Numerical Mathematics\/}~{\em 4}, 579--598.

\bibitem[\protect\citeauthoryear{Wahba, Wang, Gu, Klein, and Klein}{Wahba
  et~al.}{1995}]{wahba1995smoothing}
Wahba, G., Y.~Wang, C.~Gu, R.~Klein, and B.~Klein (1995).
\newblock Smoothing spline anova for exponential families, with application to
  the wisconsin epidemiological study of diabetic retinopathy: the 1994 neyman
  memorial lecture.
\newblock {\em Annals of Statistics\/}~{\em 23\/}(6), 1865--1895.

\bibitem[\protect\citeauthoryear{Wang and Lin}{Wang and
  Lin}{2015}]{wang2015derivative}
Wang, W.~W. and L.~Lin (2015).
\newblock Derivative estimation based on difference sequence via locally
  weighted least squares regression.
\newblock {\em Journal of Machine Learning Research\/}~{\em 16\/}(1),
  2617--2641.

\bibitem[\protect\citeauthoryear{Wang}{Wang}{1998}]{wang1998smoothing}
Wang, Y. (1998).
\newblock Smoothing spline models with correlated random errors.
\newblock {\em Journal of the American Statistical Association\/}~{\em
  93\/}(441), 341--348.

\bibitem[\protect\citeauthoryear{Wang, Zhou, Li, and Lian}{Wang
  et~al.}{2022}]{wang2022sparse}
Wang, Y., Y.~Zhou, R.~Li, and H.~Lian (2022).
\newblock Sparse high-dimensional semi-nonparametric quantile regression in a
  reproducing kernel hilbert space.
\newblock {\em Computational Statistics \& Data Analysis\/}~{\em 168}, 107388.

\bibitem[\protect\citeauthoryear{Woodall, Spitzner, Montgomery, and
  Gupta}{Woodall et~al.}{2004}]{woodall2004using}
Woodall, W.~H., D.~J. Spitzner, D.~C. Montgomery, and S.~Gupta (2004).
\newblock Using control charts to monitor process and product quality profiles.
\newblock {\em Journal of Quality Technology\/}~{\em 36\/}(3), 309--320.

\bibitem[\protect\citeauthoryear{Xia}{Xia}{1998}]{xia1998bias}
Xia, Y. (1998).
\newblock Bias-corrected confidence bands in nonparametric regression.
\newblock {\em Journal of the Royal Statistical Society: Series B (Statistical
  Methodology)\/}~{\em 60\/}(4), 797--811.

\bibitem[\protect\citeauthoryear{Yang, Sperlich, and Härdle}{Yang
  et~al.}{2003}]{YANG2003521}
Yang, L., S.~Sperlich, and W.~Härdle (2003).
\newblock Derivative estimation and testing in generalized additive models.
\newblock {\em Journal of Statistical Planning and Inference\/}~{\em 115\/}(2),
  521--542.

\bibitem[\protect\citeauthoryear{Ye and Li}{Ye and Li}{2025}]{YeLi2025}
Ye, S. and M.~Li (2025).
\newblock Ab initio nonparametric variable selection for scalable symbolic
  regression with large p.
\newblock In {\em Proceedings of the 42nd International Conference on Machine
  Learning (ICML)}, Volume 267 of {\em Proceedings of Machine Learning
  Research}, pp.\  72041--72062. PMLR.

\bibitem[\protect\citeauthoryear{Ye, Senftle, and Li}{Ye
  et~al.}{2024}]{YeSenftleLi2024}
Ye, S., T.~P. Senftle, and M.~Li (2024).
\newblock Operator-induced structural variable selection for identifying
  materials genes.
\newblock {\em Journal of the American Statistical Association\/}~{\em
  119\/}(545), 81--94.

\bibitem[\protect\citeauthoryear{Yu, Shi, Liu, and Wang}{Yu
  et~al.}{2022}]{yu2022smoothing}
Yu, J., J.~Shi, A.~Liu, and Y.~Wang (2022).
\newblock Smoothing spline semiparametric density models.
\newblock {\em Journal of the American Statistical Association\/}~{\em
  117\/}(537), 237--250.

\bibitem[\protect\citeauthoryear{Yuan and Cai}{Yuan and
  Cai}{2010}]{yuancai2010rkhs}
Yuan, M. and T.~T. Cai (2010).
\newblock {A reproducing kernel Hilbert space approach to functional linear
  regression}.
\newblock {\em Annals of Statistics\/}~{\em 38\/}(6), 3412 -- 3444.

\bibitem[\protect\citeauthoryear{Zhang, Liu, and Wu}{Zhang
  et~al.}{2016}]{zhang2016quantile}
Zhang, C., Y.~Liu, and Y.~Wu (2016).
\newblock On quantile regression in reproducing kernel hilbert spaces with the
  data sparsity constraint.
\newblock {\em The Journal of Machine Learning Research\/}~{\em 17\/}(1),
  1374--1418.

\bibitem[\protect\citeauthoryear{Zhang, Jin, Wang, Sun, Ma, and Zhong}{Zhang
  et~al.}{2018}]{zhang2018smoothing}
Zhang, J., H.~Jin, Y.~Wang, X.~Sun, P.~Ma, and W.~Zhong (2018).
\newblock Smoothing spline anova models and their applications in complex and
  massive datasets.
\newblock In Y.~K.-N. Truong and M.~Sarfraz (Eds.), {\em Topics in Splines and
  Applications}, Chapter~4. Rijeka: IntechOpen.

\bibitem[\protect\citeauthoryear{Zhang}{Zhang}{2005}]{zhang2005learning}
Zhang, T. (2005).
\newblock Learning bounds for kernel regression using effective data
  dimensionality.
\newblock {\em Neural Computation\/}~{\em 17\/}(9), 2077--2098.

\bibitem[\protect\citeauthoryear{Zhang, Duchi, and Wainwright}{Zhang
  et~al.}{2015}]{zhang2015divide}
Zhang, Y., J.~Duchi, and M.~Wainwright (2015).
\newblock Divide and conquer kernel ridge regression: A distributed algorithm
  with minimax optimal rates.
\newblock {\em Journal of Machine Learning Research\/}~{\em 16\/}(1),
  3299--3340.

\bibitem[\protect\citeauthoryear{Zhao, Liu, and Shang}{Zhao
  et~al.}{2021}]{zhao2021statistical}
Zhao, S., R.~Liu, and Z.~Shang (2021).
\newblock Statistical inference on panel data models: a kernel ridge regression
  method.
\newblock {\em Journal of Business \& Economic Statistics\/}~{\em 39\/}(1),
  325--337.

\end{thebibliography}
\clearpage

\end{document}